\documentclass [12pt]{article}
\usepackage [cp1251]{inputenc}
\usepackage [english]{babel}
\usepackage {graphicx}
\usepackage {amssymb}
\usepackage {amsmath}
\usepackage {longtable}
\title{Nonlinear dynamo in obliquely rotating electroconductive fluids}

\author{$^{1}$\textbf{M.I. Kopp}, $^3$\textbf{A.V. Tur}, $^{1,2}$\textbf{V.V. Yanovsky}}

\begin{document}

 \maketitle

$^{1}$ \textit{Institute for Single Crystals, NAS  Ukraine, Nauky Ave. 60, Kharkov 61001, Ukraine}

$^{2}$\textit{V.N. Karazin Kharkiv National University 4 Svobody Sq., Kharkov 61022, Ukraine}

$^{3}$\textit{Universit\'{e} de Toulouse [UPS], CNRS, Institut de Recherche en Astrophysique et Plan\'{e}tologie,
9 avenue du Colonel Roche, BP 44346, 31028 Toulouse Cedex 4, France}

\abstract{In the present paper, we study a new type of large-scale instability, which arises  in  obliquely rotating electroconductive fluids with a small-scale  external force of zero helicity. This force excites small-scale velocity oscillations with a small Reynolds number. We used the method of multiscale asymptotic expansions. The nonlinear equations  for vortex and magnetic perturbations motions are obtained  up to third order  in Reynolds number. The linear stage of the magneto-vortex dynamo, arising as a result of instabilities of the type of  hydrodynamic and magnetohydrodynamic  $\alpha $ - effects, is investigated. Stationary solutions of nonlinear equations of magneto-vortex dynamo in the form of localized chaotic structures are found numerically. }


\section{Introduction}
Recently, a great interest was attracted to the origin of magnetic fields in various astrophysical objects such as planets, the Sun, stars and galaxies.
By present, the direction of these studies formed itself into an independent section of physics, that is the dynamo theory \cite{1s}-\cite{10s}. It is obvious that an important role in dynamo theory is played by the phenomenon of the rotational motion of cosmic bodies. Due to this rotation, various wave  (Rossby waves, inertial waves)  and vortex motions (geostrophic vortices, etc.) are excited \cite{11s}-\cite{18s}. This way, under the influence of the Coriolis force, initially  mirror-symmetric  turbulence becomes spiral, which is characterized by  violation of mirror symmetry of the turbulent fluid motion. In \cite{19s}, the important topological characteristics of helical turbulence $J_s= \overline{\vec v rot\vec v}$  was introduced, that   is a measure of the engagement force lines of the vortex field. A rigorous physical and mathematical justification for the relationship between the cross magnetic helicity $\mathcal E=\overline{\vec v \vec H}$  (or  the turbulent e.m.f.)  with the topological invariant  $J_s$  was given in \cite{20s}, where it was shown that  generation of a large-scale field occurs under the action of a turbulent e.m.f. proportional to the mean magnetic field $\vec{ {\mathcal E}}=\alpha \overline{\vec{H}}$ .  Coefficient   $\alpha $    is proportional to the mean helicity of the velocity field  $\alpha \sim \overline{\vec{v}rot\vec{v}}$   and  has got a definition of $\alpha $ - effect. On the basis of  $\alpha $ - effect, various theories have been constructed that explain the origin of magnetic fields in various astrophysical objects: the planets and the sun\cite{1s}-\cite{6s} , galaxies \cite{7s}  and so on. A recent review \cite{8s} discusses in detail the issues of laboratory modeling of $\alpha $ - effect.
 Generation properties of spiral turbulence were discovered not only in magnetic hydrodynamics or electrically conductive media, but also in  hydrodynamics. The hypothesis that helical turbulence can generate large eddies was first suggested in \cite{21s}. This  hypothesis was based on the formal similarity of the equations of the magnetic field  $\vec{H}$   and the equation for the vorticity   $\vec{\omega }=rot\vec{v}$. However, in \cite{22s} the proof was given cited evidence that, in  incompressible turbulent fluid, $\alpha $ - effect is not possible due to the certain symmetry of the tensor of Reynolds stresses in the averaged Navier-Stokes equation. Thus, sole helicity turbulence is not sufficient for the appearance of the hydrodynamic (HD) $\alpha $ - effect. Other factors  are needed of symmetry breaking of the turbulent flow. As it is shown in \cite{23s}, one of  such factors is the compressibility of the medium, and in  \cite{24s} - temperature gradient in the gravity field. The effect of generation of  large scale vortex structures (LSVS) by  spiral turbulence has been called the vortex dynamo.  The vortex dynamo mechanisms were developed with reference to the turbulent atmosphere and the ocean. The theory of the convective vortex dynamo was  substantionally developed in \cite{24s}-\cite{31s}, where spiral turbulence led to large-scale instability. Because of this instability, one convective cell is formed, which is  interpreted as a huge vortex such as a tropical cyclone.
There is also a large number of articles  \cite{32s}-\cite{36s}  on the generation of  LSVS with allowance for the rotation effects.  A fundamentally different  $\alpha $-- effect was discovered in \cite{37s}, where the turbulent fluid motion is simulated by the external small-scale force $\vec{F}_{0} $.  The model of the external small-scale force was chosen with parity violation (for zero helicity $\vec{F}_{0} rot\vec{F}_{0} =0$ ). The effect of the generation of large-scale perturbations by such a force is called the anisotropic kinetic alpha-effect or the AKA - effect \cite{36s}.  In the present paper, a large-scale instability in an incompressible fluid were considered by the method of  asymptotic multiscale expansions.   The Reynolds number  $R=\frac{v_{0} t_{0} }{\lambda _{0} } \ll 1$ is used as a small parameter for the asymptotic method of multiscale expansions for small-scale velocity fluctuations  $v_{0}$ caused by small-scale force. The application of the method of multiscale asymptotic expansions allowed the development of linear and nonlinear vortex dynamo theories for compressible media \cite{38s}, \cite{39s} as well as for   convective media with spiral external force \cite{29s}-\cite{31s}.
In the above-mentioned articles, spiral turbulence is  assumed to be known, or the problem  of its generation was considered independently \cite{40s}. Obviously, the question arises of the possibility of generation of large-scale  fields (vortex and magnetic)  in a rotating media under influence of small-scale force with zero helicity  $\vec{F}_{0} rot\vec{F}_{0} =0$ .  An example of the generation of LSVS in a rotating incompressible fluid was found in \cite{41s}. It was also shown that the development of large-scale instability in an inclined rotating fluid generates  nonlinear large-scale helical vortex structures or localized Beltrami vortices with the internal helical structure.
In the present paper, we give a generalization of  $\alpha $ - effect found in \cite{41s} for the case of electroconductive  fluid. As a result of this generalization, we obtain the large-scale instability which generates the LSVS and large-scale  magnetic fields.
 The organization of this paper is as follows. In Sec. 2 are set out the basic equations and the formulation of the problem. In the Sec. 3 we obtain the averaged magnetohydrodynamic equations in an obliquely rotating fluid for the large-scale fields using the method of multiscale asymptotic developments. The technical aspect of this question is described in detail in Appendix I. The correlation functions in the averaged equations are expressed in terms of  small-scale fields in  zero-order approximation with respect to $R$. In Appendix II in order to obtain the averaged equations in closed form, we find solutions for small-scale fields in zero order approximation. In Appendix III we calculate the Reynolds stresses, Maxwell stresses and turbulent e.m.f. using these solutions. As a result, in Sec. 3 self-consistent system of nonlinear equations is obtained for large scale magnetic and hydrodynamic fields. Unlike kinematic dynamo, these fields have mutual influence on each other. A system of equations describing the mutual influence of fields is called the equations of a nonlinear magnetic-vortex dynamo. In Sec. 4, the stability of small large-scale vortex and magnetic disturbances is investigated. In Sec. 5, a numerical analysis of the nonlinear equations in the steady-state regime is carried out.  There is also demonstrated  the existence of chaotic localized vortex and magnetic structures.
The results can be applied to a number of astrophysical problems.

\section{Basic equations and formulation of the problem}

\begin{figure}
  \centering
  \includegraphics[width=6 cm]{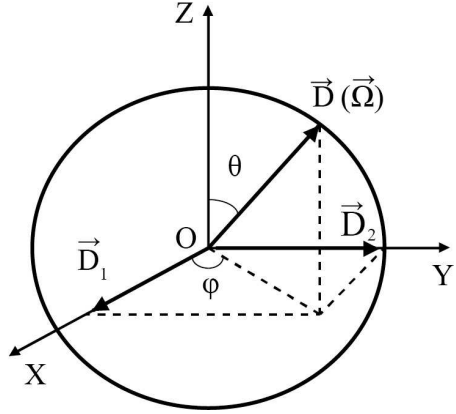}\\
  \caption{ In general, the angular  $\vec{\Omega}$ is inclined with respect to the plane $(X,Y)$  of   the external force  $ \vec{F}_{0 \bot }$  is located. A connection is shown between the Cartesian projections of the rotational parameter   $\vec D$   (or the angular velocity vector of rotation  $\vec \Omega$) with their projections in a spherical coordinate system.}\label{fg1}
\end{figure}

The starting equations for describing the dynamics of a rotating electrically conducting incompressible fluid are the well-known equations of magnetohydrodynamics:
\begin{equation} \label{eq1}
 \frac{\partial \vec{V}}{\partial t} +\left(
\vec{V}\nabla \right)\vec{V}=-\frac{\nabla P}{\rho _{00} } +2\left[\vec{V}
\times \vec{\Omega}\right]+\frac{1}{4\pi \rho _{00} } \left[rot\vec{B}\times
\vec{B}\right]+\nu \Delta \vec{V}+\vec{F}_{0}
\end{equation}

\begin{equation} \label{eq2}
 \frac{\partial \vec{B}}{\partial t} =rot\left[
 \vec{V}\times \vec{B}\right]+\nu _{m} \Delta \vec{B}
\end{equation}

\begin{equation}\label{eq3}
  div \vec{V}=0 \; \;\;div \vec{B}=0
\end{equation}
Here  $\vec{V}$, $P$, $\vec{B}$ are correspondent   perturbations of velocity, pressure and magnetic field relative to the equilibrium state:
\begin{equation} \label{eq4}
 \nabla P_{00} =-\rho _{00} \nabla {\Phi}_{00} -\rho _{00} \left[\vec{\Omega}\times \left[\vec{\Omega}\times
\vec{r}\right]\right]
\end{equation}
where  $\vec{r}$  is a radius-vector of an element of the environment.  $\Phi_{00}$ - the equilibrium potential corresponding to the external force of gravity,  $\rho _{00}$ - the equilibrium medium density:  $\rho _{00} =const$ ,  $\nu $ - the coefficient of viscosity of liquid,  $\nu_{m} =\frac{c^{2} }{4\pi \sigma _{c} } $ - the coefficient of magnetic viscosity,  $\sigma_{c}$  - the conductivity of the medium. The vector of angular speed    $\vec{\Omega}$   we consider to be constant (solid rotation) and inclined relative to the plane  $(X,Y)$ , as shown in Fig. \ref{fg1}, i.e., for a Cartesian geometry problem:  $\vec{\Omega}=\left( \Omega_{1}, \Omega_{2}, \Omega_{3}\right)$. In addition,  we assume that the medium is unbounded and we disregard the effect of the external magnetic field.  In this case, small-scale magnetic fields or so-called seed magnetic fields can be excited  by not  turbulent mechanisms, for example as a result of the development of hydrodynamic instabilities \cite{1s}-\cite{3s}, thermomagnetic instabilities \cite{42s} etc. This formulation of the problem is of interest for the dynamo theory \cite{1s}-\cite{10s}. The external force $\vec{F}_{0}$  is included in the equation (\ref{eq1}), that simulates the excitation source in a medium of small-scale  high-frequency fluctuations of the velocity field   ${\vec{v}}_0$  with a small Reynolds number  $R=\frac{v_{0} t_{0} }{\lambda _{0} } \ll 1$. In contrast to the previous studies \cite{29s}-\cite{31s}, \cite{38s}, \cite{39s}, here we   consider   the non-helical outer
force  ${\vec{F}}_0$   with the following properties:
\begin{equation} \label{eq5}
div\vec{F}_{0} =0, \; \vec{F}_{0} rot\vec{F}_{0}
=0,\;  rot\vec{F}_{0} \ne 0 , \;    \vec{F}_{0} =f_{0} \vec{F}_{0} \left(\frac{x}{\lambda
_{0} } ;  \frac{t}{t_{0} } \right)
\end{equation}
where  $\lambda _{0}$  is the characteristic scale, $t_{0}$   is the characteristic time,  $f_{0}$  is the characteristic amplitude. The external force is set in the plane  $(X,Y)$  where normal  projection of the angular velocity $\vec{\Omega }$  lies. We choose an external force in a rotating coordinate system in the following form:
\begin{equation} \label{eq6}
 \begin{array}{l} {F_{0} ^{z} =0,\;\vec{F}_{0
\bot } =f_{0} \left(\vec{i}cos\phi _{2} +\vec{j}cos\phi _{1}
\right),\; \phi _{1} =\vec{k}_{1} \vec{x}-\omega _{0} t,\phi _{2}
=\vec{k}_{2} \vec{x}-\omega _{0} t,} \\ {\vec{k}_{1}
=k_{0} \left(1,0,0\right),\vec{k}_{2} =k_{0} \left(0,1,0\right).}
\end{array}
 \end{equation}

Obviously, this external non-spiral force satisfies all the conditions (\ref{eq5}). Let us turn in equations (\ref{eq1})-(\ref{eq3}) to dimensionless variables. For the sake of convenience of notation, let the notations of dimensional variables are preserved:
\[\vec{x}
\to \frac{\vec{x}}{\lambda _{0} } ,\quad t\to \frac{t}{t_{0} } ,\; \; \;
\; \vec{V}\to \frac{\vec{V}}{v_{0} } ,\quad \vec{F}_{0} \to \frac{\vec{F}_{0}
}{f_{0} } ,\quad \vec{B}\to \frac{\vec{B}}{B_{0} } ,\quad t_{0} =\frac{\lambda _{0}^{2}
}{\nu } , \]
\[ f_{0} =\frac{v_{0} \nu }{\lambda _{0}^{2} } \quad P\to \frac{P}{P_{0}
\rho _{00} } ,\quad  P_{0} =\frac{\nu v_{0} }{\lambda _{0} }\]
In dimensionless variables, the equations (\ref{eq1})-(\ref{eq3}) take the form:
\begin{equation} \label{eq7}
\frac{\partial V_{i} }{\partial t} +RV_{k} \frac{
\partial V_{i} }{\partial x_{k} } =-\frac{1}{\rho _{00} } \frac{\partial P}{\partial
x_{i} } +\varepsilon _{ijk} V_{j} D_{k} +\widetilde{Q}R\varepsilon _{ijk} \varepsilon
_{jml} \frac{\partial B_{l} }{\partial x_{m} } B_{k} +\frac{\partial ^{2} V_{i} }{
 \partial x_{k}^{2} } +F_{0}^{i}
\end{equation}

\begin{equation} \label{eq8}
  \frac{\partial B_{i} }{\partial t} -Pm^{-1}
\frac{\partial ^{2} B_{i} }{\partial x_{k}^{2} } =R\varepsilon _{ijk} \varepsilon
_{knp} V_{n} B_{p}
\end{equation}

\begin{equation} \label{eq9}
 \frac{\partial V_{k} }{\partial x_{k} } =\frac{
\partial B_{k} }{\partial x_{k} } =0
\end{equation}
The nature of the evolution of the fields described by equation system (\ref{eq7})-(\ref{eq9}) is determined to a large extent by the following dimensionless parameters: $D_{i}=\frac{2{\Omega}_{i} \lambda _{0}^{2} }{\nu } $  is the dimensionless rotational parameter on the scale $\lambda _{0}$ ($i=1,2,3$) associated with the Taylor number $Ta_{i} =D_{i}^{2} $ \cite{11s}-\cite{13s} and is a characterizes of the degree of influence of Coriolis forces over viscous forces;  $\widetilde{Q}=\frac{Q}{Pm} $, $Q=\frac{\sigma _{c} B_{0}^{2} \lambda _{0}^{2}}{c^{2} \rho _{00} \nu } $ is the Chandrasekhar number, $Pm=\frac{\nu }{\nu _{m} } $ is the magnetic Prandtl number. A small parameter of the asymptotic expansion is the Reynolds number $R=
\frac{v_{0} t_{0} }{\lambda _{0} } \ll 1$. The parameters   $D$  and   $\widetilde{Q}$  are arbitrary not affecting the decomposition scheme. This is the external force, acting at the equilibrium state, that causes small-scale  high-frequency oscillations of the velocity. The mean values of such oscillations are zero, but because of the nonlinear interaction in certain orders of perturbation theory there arise terms that do not vanish upon averaging. Such terms are called secular and they will be a condition for the solvability of a multiscale asymptotic expansion. Finding the solvability equations, which determine the evolution of large-scale perturbations  is the main task.

\section{Equations of a nonlinear magnetic-vortex dynamo in a quasi-two-dimensional model}

Let us consider in more detail the application of the method of multiscale asymptotic expansions to the problem of nonlinear evolution of large-scale vortex and magnetic perturbations in an obliquely rotating electrically conductive medium. The method of constructing asymptotic equations is well developed in the papers \cite{37s}, \cite{29s}-\cite{31s}, following which we represent  spatial and temporal derivatives in the equations (\ref{eq7})-(\ref{eq9}) in the form of asymptotic expansions:
\begin{equation} \label{eq10}
 \frac{\partial }{\partial t} \to \partial _{t}
+R^{4} \partial _{T}, \;\;\; \frac{\partial }{\partial x_{i} } \to \partial _{i} +R^{2}
\nabla _{i}
\end{equation}
where   $\partial _{i} $  and  $\partial _{t} $  denote the derivatives with respect to the fast variables  $x_{0}=\left(\vec{x}_{0} ,t_{0} \right)$,  while  $\nabla _{i} $ and   $\partial _{T} $  are derivatives with respect to slow variables $X=\left(\vec{X},\; T\right)$ . $x_{0}$   and   $X$  can be called, respectively, small-scale and large-scale variables. When constructing a nonlinear  theory, the variables $\vec{V}$, $\vec{B}$, $P$ can be represented as asymptotic series:
\[\vec{V}
\left(\vec{x},t\right)\; \; =\frac{1}{R} \vec{W}_{-1} \left(X\right)+
\vec{v}_{0} \left(x_{0} \right)+R\vec{v}_{1} +R^{2} \vec{v}_{2} +R^{3} \vec{v}_{3}
+\cdots  \]

\begin{equation} \label{eq11}
 \vec{B}\left(\vec{x},t\right)=\frac{1}{R} \vec{B}_{-1}
\left(X\right)+\vec{B}_{0} \left(x_{0} \right)+R\vec{B}_{1} +R^{2} \vec{B}_{2} +R^{3}
\vec{B}_{3} +\cdots
\end{equation}

\[P(x)=\frac{1}{R^{3} } P_{-3} +\frac{1}{R^{2} } P_{-2} +\frac{1}{R} P_{-1} +P_{0}
+R(P_{1} +\overline{P}_{1} \left(X\right))+R^{2} P_{2} +R^{3} P_{3} +\cdots \]
Substituting the expansions (\ref{eq10})-(\ref{eq11}) into equation system (\ref{eq7})-(\ref{eq9}) and collecting together the terms with the same orders of $R$ up to the degree $R^3$  inclusive, we obtain the equations of a multiscale asymptotic expansion. The algebraic structure of the asymptotic expansion of the equations (\ref{eq7})-(\ref{eq9})  with different orders of  R  is given in Appendix I. It also shown there, that, up to $R^3$,  the basic secular equations are obtained,  i.e. equations for large-scale fields:
\begin{equation} \label{eq12}
  \partial _{t} W_{-1}^{i} -\Delta W_{-1}^{i}
+{{\nabla }_{k}}\overline{(v_{0}^{k}v_{0}^{i})}=-\nabla _{i} \overline{P}_{1}
+\widetilde{Q}\varepsilon _{ijk} \varepsilon _{jml} \left({{\nabla }_{m}}\overline{(B_{0}^{l}B_{0}^{k})}\right)
\end{equation}

\begin{equation} \label{eq13}
  \partial _{t} B_{-1}^{i} -Pm^{-1} \Delta B_{-1}^{i}
=\varepsilon _{ijk} \varepsilon _{knp} {{\nabla }_{j}}\overline{(v_{0}^{n}B_{0}^{p})}
\end{equation}
Using the convolution of tensors $\varepsilon _{ijk} \varepsilon _{jml} =\delta _{km} \delta
_{il} -\delta _{im} \delta _{kl} ,\; $$\varepsilon _{ijk} \varepsilon _{knp} =\delta
_{in} \delta _{jp} -\delta _{ip} \delta _{jp} $,   and introducing the notation   $\vec{W}=
\vec{W}_{-1} ,\;  \vec{H}=\vec{B}_{-1} $, we reorganize the equation system (\ref{eq12})-(\ref{eq13})  into the following form:
\begin{equation} \label{eq14}
   \partial _{T} W_{i} -\Delta W_{i} +{{\nabla }_{k}}\overline{(v_{0}^{k}v_{0}^{i})}=-\nabla _{i} \overline{P}_{1} +\widetilde{Q}\left( {{\nabla }_{k}}\overline{(B_{0}^{i}B_{0}^{k})}-\frac{{{\nabla }_{i}}}{2}\overline{{{(B_{0}^{k})}^{2}}} \right)
\end{equation}

\begin{equation} \label{eq15}
  \partial _{T} H_{i} -Pm^{-1} \Delta H_{i} =
{{\nabla }_{j}}\overline{(v_{0}^{i}B_{0}^{j})}-{{\nabla }_{j}}\overline{(v_{0}^{j}B_{0}^{i})}
\end{equation}
The equations (\ref{eq14})-(\ref{eq15}) are supplemented by secular equations that were obtained in Appendix I:

\[\nabla _{k} \left(W_{k} W_{i} \right)=-\nabla
_{i} \ P_{-1} +\widetilde{Q}\left(\nabla _{k} H_{i} -\nabla _{i} { H}_{{
 k}} \right){ H}_{{ k}} \]

\[ W_{j} \nabla _{j} H_{i} =H_{j} \nabla _{j} W_{i}  \]

\[ W_{j} \nabla _{j} H_{i} =H_{j} \nabla _{j}W_{i} \]

\[ \nabla _{i} W_{i} =0, \;\nabla _{i} H_{i} =0,\;   \nabla _{i} P_{-3} =\varepsilon _{ijk} W_{j} D_{k} \]

Thus,  in order  to obtain the equation system \eqref{eq14}-\eqref{eq15} describing the evolution of large-scale  $\vec{W}$ and $\vec{H}$, one must reach the third order of perturbation theory. This is a rather characteristic phenomenon when applying the method of multiscale expansions. The equations \eqref{eq14}-\eqref{eq15} acquire a closed form after the calculation of correlation functions: Reynolds stresses ${{\nabla }_{k}}\overline{(v_{0}^{k}v_{0}^{i})}$ ,  Maxwell stresses  ${{\nabla }_{k}}\overline{(B_{0}^{i}B_{0}^{k})}$ and  turbulent e.m.f. ${{\mathcal{E}}_{n}}={{\varepsilon }_{nij}}\overline{v_{0}^{i}B_{0}^{j}}$. The calculation of these correlation functions is greatly simplified if we use the quasi-two-dimensional approximation, which is employed to describe large-scale vortex and magnetic fields in many astrophysical and geophysical problems \cite{3s,14s,30s,31s}. In the framework of this approximation, for our problem, we  assume that the large-scale derivative with respect to $Z$ is more preferable, i.e.
\[\qquad \frac{\partial }{\partial  Z} \gg \frac{\partial }{\partial X} , \frac{\partial }{\partial Y}, \]
and the geometry of large-scale fields has the following form:
\begin{equation} \label{eq16}
   \vec{W}=\left(W_{1} \left(Z\right),\; W_{2}
\left(Z\right),\; 0\right), \vec{H}=\left(H_{1}\left(Z\right),\; H_{2}\left(Z\right),
\; 0\right)
\end{equation}
This geometry corresponds to large-scale fields of the Beltrami type:   $\vec{W}\times rot\vec{W}=0$   and  $\vec{H}\times rot\vec{H}=0$. Within the quasi-two-dimensional  problem, the equation system (\ref{eq14})-(\ref{eq15}) is simplified:
\begin{equation} \label{eq17} \partial _{T} W_{1} -\Delta W_{1} +{{\nabla }_{Z}}\overline{(v_{0}^{z}v_{0}^{x})}=\widetilde{Q}{{\nabla }_{Z}}\overline{(B_{0}^{z}B_{0}^{x})}\end{equation}

\begin{equation} \label{eq18} \partial _{T} W_{2} -\Delta W_{2} +{{\nabla }_{Z}}\overline{(v_{0}^{z}v_{0}^{y})}=\widetilde{Q}{{\nabla }_{Z}}\overline{(B_{0}^{z}B_{0}^{y})}\end{equation}

\begin{equation} \label{eq19} \partial _{T} H_{1} -Pm^{-1} \Delta H_{1} =
{{\nabla }_{Z}}\overline{(v_{0}^{x}B_{0}^{z})}-{{\nabla }_{Z}}\overline{(v_{0}^{z}B_{0}^{x})} \end{equation}

\begin{equation} \label{eq20} \partial _{T} H_{2} -Pm^{-1} \Delta H_{2} ={{\nabla }_{Z}}\overline{(v_{0}^{y}B_{0}^{z})}-{{\nabla }_{Z}}\overline{(v_{0}^{z}B_{0}^{y})}
 \end{equation}
To obtain the equations \eqref{eq17}-\eqref{eq20} in the closed form, we use solutions of the equations for small-scale fields at zero order in  $R$, obtained in Appendix II. Next,  it is necessary to calculate the correlators that are included into  the equation system \eqref{eq17}-\eqref{eq20}:
\[ T^{31} =\overline{w_{0} u_{0} }=\overline{w_{01} \left(u_{01}
\right)^{*} }+\overline{\left(w_{01} \right)^{*} u_{01} }+\overline{w_{03} \left(u_{03}
\right)^{*} }+\overline{\left(w_{03} \right)^{*} u_{03} } \]

 \[ T^{32} =\overline{w_{0} v_{0} }=\overline{w_{01}
\left(v_{01} \right)^{*} }+\overline{\left(w_{01} \right)^{*} v_{01} }+\overline{w_{03}
\left(v_{03} \right)^{*} }+\overline{\left(w_{03} \right)^{*} v_{03} } \]

 \[ S^{31} =\overline{\widetilde{w}_{0} \widetilde{u}_{0} }=\overline{
\widetilde{w}_{01} \left(\widetilde{u}_{01} \right)^{*} }+\overline{\left(\widetilde{w}_{01}
\right)^{*} \widetilde{u}_{01} }+\overline{\widetilde{w}_{03} \left(\widetilde{u}_{03}
\right)^{*} }+\overline{\left(\widetilde{w}_{03} \right)^{*} \widetilde{u}_{03} } \]

  \[ S^{32} =\overline{\widetilde{w}_{0} \widetilde{v}_{0} }=\overline{
\widetilde{w}_{01} \left(\widetilde{v}_{01} \right)^{*} }+\overline{\left(\widetilde{w}_{01}
\right)^{*} \widetilde{v}_{01} }+\overline{\widetilde{w}_{03} \left(\widetilde{v}_{03}
\right)^{*} }+\overline{\left(\widetilde{w}_{03} \right)^{*} \widetilde{v}_{03} } \]

 \[  G^{13} =\overline{u_{0} \widetilde{w}_{0} }=\overline{u_{01}
\left(\widetilde{w}_{01} \right)^{*} }+\overline{\left(u_{01} \right)^{*} \widetilde{w}_{01}
}+\overline{u_{03} \left(\widetilde{w}_{03} \right)^{*} }+\overline{\left(u_{03}
\right)^{*} \widetilde{w}_{03} } \]

 \[   G^{31} =\overline{w_{0} \widetilde{u}_{0} }=
\overline{w_{01} \left(\widetilde{u}_{01} \right)^{*} }+\overline{\left(w_{01} \right)^{*}
\widetilde{u}_{01} }+\overline{w_{03} \left(\widetilde{u}_{03} \right)^{*} }+\overline{
\left(w_{03} \right)^{*} \widetilde{u}_{03} } \]

 \[  G^{23} =\overline{v_{0} \widetilde{w}_{0} }=
\overline{v_{01} \left(\widetilde{w}_{01} \right)^{*} }+\overline{\left(v_{01} \right)^{*}
\widetilde{w}_{01} }+\overline{v_{03} \left(\widetilde{w}_{03} \right)^{*} }+\overline{
\left(v_{03} \right)^{*} \widetilde{w}_{03} } \]

 \[ G^{32} =\overline{w_{0} \widetilde{v}_{0} }=
\overline{w_{01} \left(\widetilde{v}_{01} \right)^{*} }+\overline{\left(w_{01} \right)^{*}
\widetilde{v}_{01} }+\overline{w_{03} \left(\widetilde{v}_{03} \right)^{*} }+\overline{
\left(w_{03} \right)^{*} \widetilde{v}_{03} } \]
Here, for convenience, new designations for small-scale fields have been adopted: $v_{0}^{x} =u_{0} $,\; $v_{0}^{y} =v_{0} $,\;$v_{0}^{z} =w_{0} $,\; $B_{0}^{x} =\widetilde{u}_{0}$,\; $B_{0}^{y} =\widetilde{v}_{0}$,\; $ B_{0}^{z} =\widetilde{w}_{0}$.   The technical side of this problem  is described in detail in Appendix III. As a result of the calculations carried out there, we obtained closed equations for large-scale fields of the velocity  $(W_1,W_2)$   and the magnetic field  $(H_1,H_2)$ in the following form
\begin{equation} \label{eq21}
 \partial _{T} W_{1} -\Delta W_{1} +\nabla _{Z}
\left(\alpha _{\left(2\right)}\cdot \left(1-W_{2} \right)\right)=0
\end{equation}

\begin{equation}\label{eq22}
 \partial _{T} W_{2} -\Delta W_{2} -\nabla _{Z} \left(\alpha
_{\left(1\right)}\cdot \left(1-W_{1} \right)\right)=0
\end{equation}

\begin{equation} \label{eq23}
 \partial _{T} H_{1} -Pm^{-1} \Delta H_{1} +
\nabla _{Z} \left(\alpha _{H}^{\left(2\right)}\cdot H_{2} \right)=0
\end{equation}

\begin{equation} \label{eq24}
 \partial _{T} H_{2} -Pm^{-1} \Delta H_{2} -
\nabla _{Z} \left(\alpha _{H}^{\left(1\right)}\cdot H_{1} \right)=0
\end{equation}
where nonlinear coefficients $\alpha _{\left(1\right)} $, $\alpha _{\left(2\right)}$, $\alpha _{H}^{\left(1\right)}$, $\alpha _{H}^{\left(2\right)}$ have the following form:
\[\alpha _{\left(1\right)} =\frac{f_{0}^{2} }{2} \frac{D_{1} q_{1} Q_{1} \left(1-W_{1} \right)^{-1} }{\left[4
\left(1-W_{1} \right)^{2} q_{1}^{2} Q_{1}^{2} +\left[D_{1}^{2} +W_{1} \left(2-W_{1}
\right)+\mu _{1} \right]^{2} \right]}\]

\[\alpha _{\left(2\right)} =\frac{f_{0}^{2} }{2} \frac{D_{2} q_{2} Q_{2} \left(1-W_{2} \right)^{-1} }{\left[4
\left(1-W_{2} \right)^{2} q_{2}^{2} Q_{2}^{2} +\left[D_{2}^{2} +W_{2} \left(2-W_{2}
\right)+\mu _{2} \right]^{2} \right]}\]

\[\alpha _{H}^{\left(1\right)}=f_{0}^{2} \frac{D_{1} \left(1-W_{1}
\right)PmQ_{1} }{\left(1+Pm^{2} \left(1-W_{1} \right)^{2} \right)\left[4\left(1-W_{1}
\right)^{2} q_{1}^{2} Q_{1}^{2} +\left[D_{1}^{2} +W_{1} \left(2-W_{1} \right)+\mu
_{1} \right]^{2} \right]}\]

\[\alpha _{H}^{\left(2\right)}=f_{0}^{2} \frac{D_{2}
\left(1-W_{2} \right)PmQ_{2} }{\left(1+Pm^{2} \left(1-W_{2} \right)^{2} \right)
\left[4\left(1-W_{2} \right)^{2} q_{2}^{2} Q_{2}^{2} +\left[D_{2}^{2} +W_{2} \left(2-W_{2}
\right)+\mu _{2} \right]^{2} \right]}\]
Here the following notation is introduced:
\[q_{1,2} =1 + \frac{Q H_{1,2}^{2}}{1+Pm^{2} \left(1-W_{1,2} \right)^{2}}, \quad Q_{1,2} =1 - \frac{QPm H_{1,2}^{2}}{1+Pm^{2} \left(1-W_{1,2} \right)^{2}},\]

\[\mu _{1,2} =(q_{1,2} -1)[2(1+Pm\left(1-W_{1,2} \right)^{2}) +(q_{1,2} -1)(1-Pm^{2}
\left(1-W_{1,2} \right)^{2} ]\]
The coefficients   $\alpha _{\left(1\right)} $, $\alpha _{\left(2\right)}$ correspond to the nonlinear HD  $\alpha $--effect  and    $\alpha _{H}^{\left(1\right)} $, $\alpha _{H}^{\left(2\right)} $ to nonlinear MHD  $\alpha $--effect.
Thus, we have obtained a self-consistent system of nonlinear evolution equations for large-scale velocity and magnetic field perturbations, which we will henceforth refer to as the equations of a nonlinear magnetic-vortex dynamo in an obliquely rotating electrically conducting liquid with a small-scale non-spiral force. If the rotation effect disappears $(\Omega =0)$, then  usual diffusion spreading of large-scale fields occurs. In the limit of a non-electrically conducting liquid  $\sigma =0$  the equations  (\ref{eq21}), (\ref{eq22}) completely coincide with the results of  \cite{41s}. Next, we consider the stability of small perturbations of fields (linear theory), and then we investigate the question of  possible existence of stationary structures.
\begin{figure}
  \centering
	\includegraphics[width=7 cm]{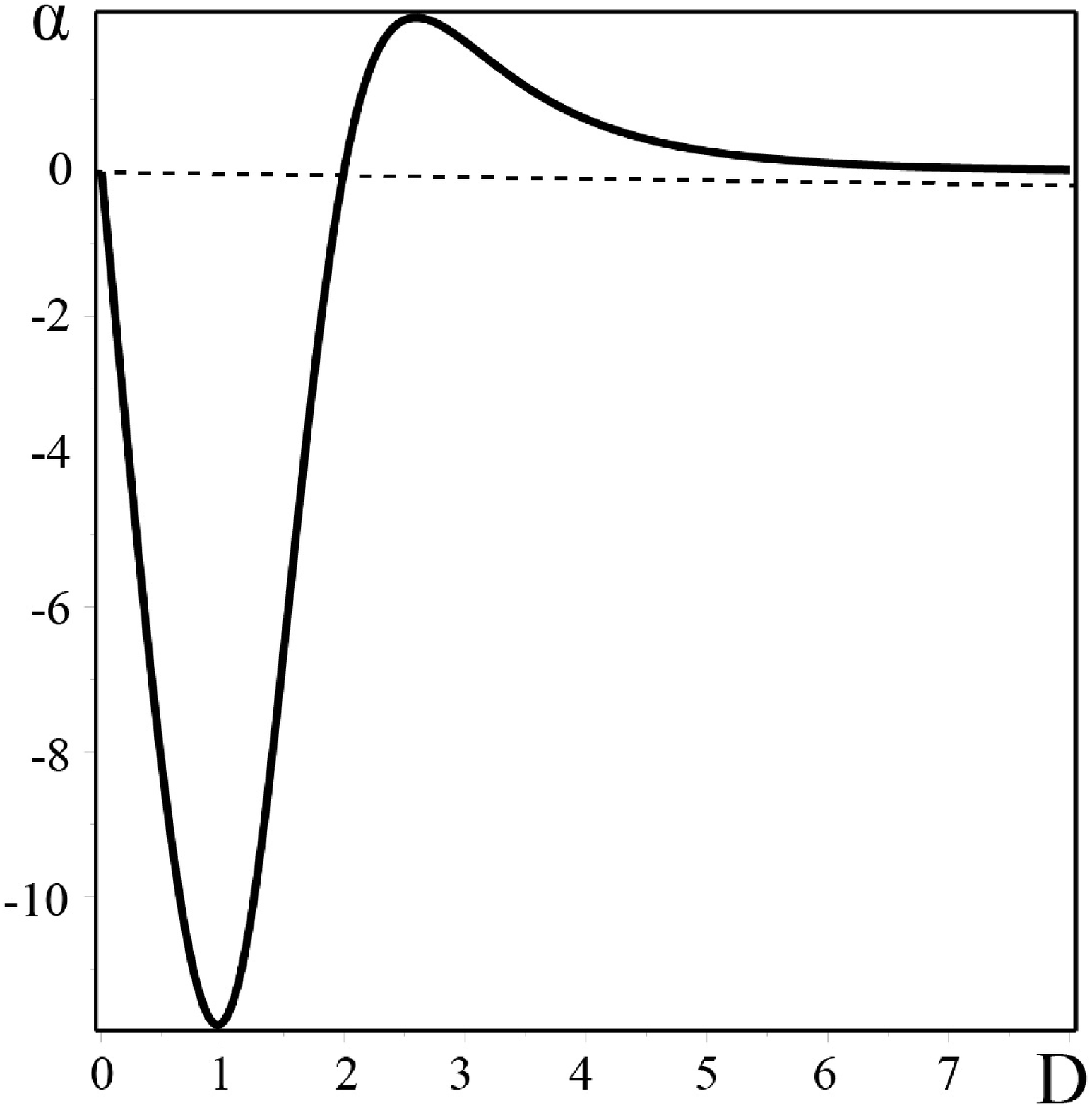}
  \includegraphics[width=7 cm]{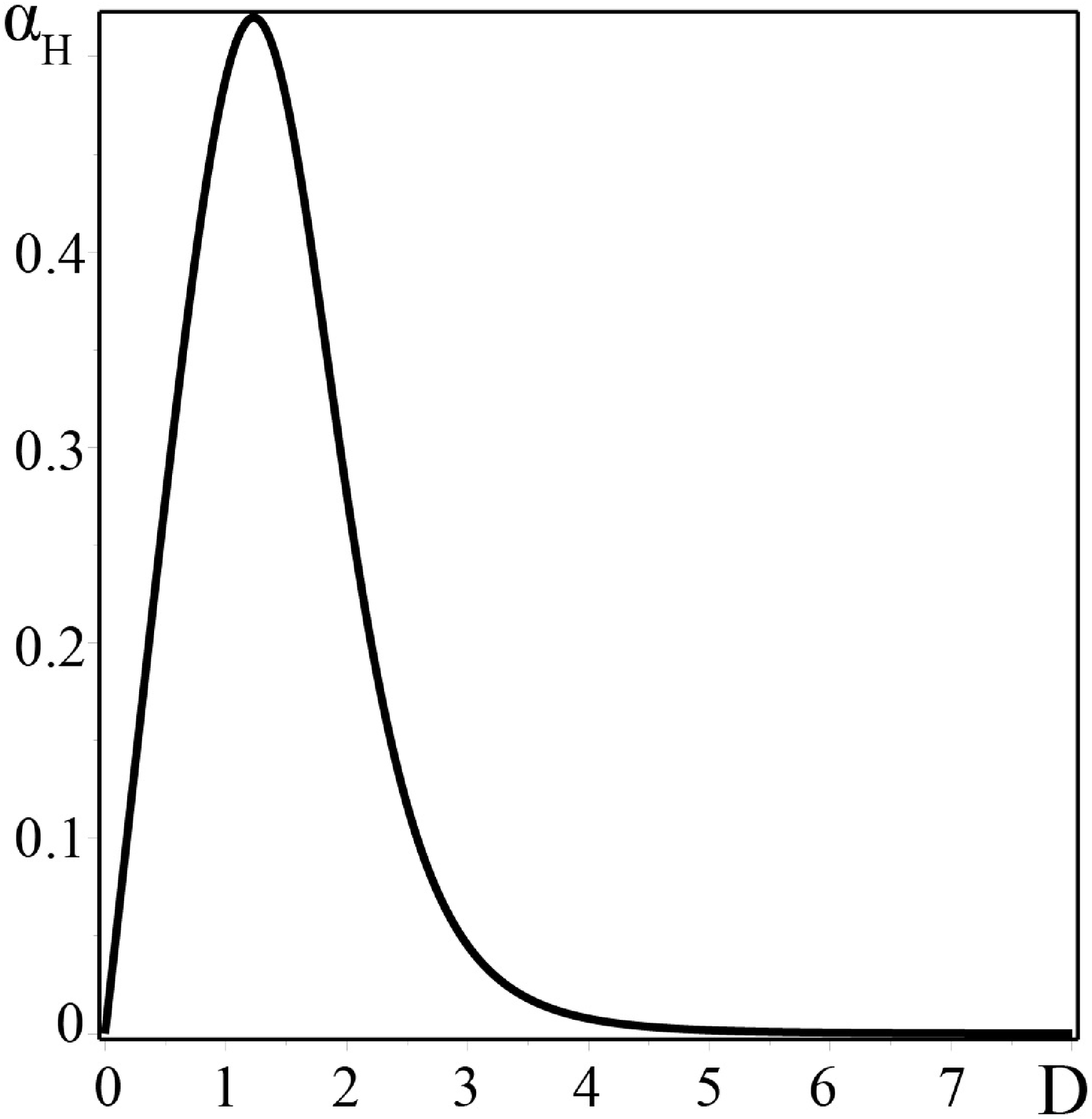}\\
  \caption{On the left is a plot of the dependence of  HD $\alpha $--effect on the parameter $D$ of fluid rotation. On the right is a plot of the dependence of  MHD $\alpha $--effect on the parameter $D$ of fluid rotation. The plots are constructed for  $\theta =\pi /2$ and  $f_{0}=10$.}\label{fg2}
\end{figure}
	
\section{Large-scale instability}

Consider the initial stage of development of the perturbations   $\left(W_{1}, W_{2} \right)$  and  $\left(H_{1}, H_{2} \right)$. Then for small values of   $\left(W_{1}, W_{2} \right)$   and  $\left(H_{1}, H_{2} \right)$   equations (\ref{eq21})-(\ref{eq24}) are linearized and reduce to the following system of  linear equations:
\begin{equation} \label{eq25}
 \left\{\begin{array}{c} {\partial _{T} W_{1}
-\nabla _{Z}^{2} W_{1} -\alpha _{2} \nabla _{Z} W_{2} =0} \\ {\partial _{T} W_{2}
-\nabla _{Z}^{2} W_{2} +\alpha _{1} \nabla _{Z} W_{1} =0} \end{array}\right.
\end{equation}

\begin{equation}\label{eq26}
 \left\{\begin{array}{c} {\partial _{T} H_{1} -Pm^{-1} \nabla
_{Z}^{2} H_{1} +\alpha _{H} ^{(2)} \nabla _{Z} H_{2} =0} \\ {\partial _{T} H_{2}
-Pm^{-1} \nabla _{Z}^{2} H_{2} -\alpha _{H} ^{(1)} \nabla _{Z} H_{1} =0} \end{array}
\right.
\end{equation}
where we have introduced the following notation for the coefficients:
\begin{equation} \label{eq27}
 \alpha _{1} =f_{0}^{2} \frac{2D_{1} \left(D_{1}^{2}
-2\right)}{\left(D_{1}^{4} +4\right)^{2} } , \;\;   \alpha _{2} =f_{0}^{2} \frac{2D_{2}
\left(D_{2}^{2} -2\right)}{\left(D_{2}^{4} +4\right)^{2} }
\end{equation}

\begin{equation} \label{eq28}
 \alpha _{H}^{(1)} =f_{0}^{2} \frac{PmD_{1}
}{\left(1+Pm^{2} \right)\left(D_{1}^{4} +4\right)},\;\;   \alpha _{H}^{(2)} =f_{0}^{2}
\frac{PmD_{2} }{\left(1+Pm^{2} \right)\left(D_{2}^{4} +4\right)}
\end{equation}
As can be seen from equation system (\ref{eq25}), (\ref{eq26}) under a small perturbation of fields  the splitting occurs  of the self-consistent system of equations (\ref{eq21})-(\ref{eq24}) into two pairs of equations for the large-scale velocity field $\vec{W}$   and magnetic field $\vec{H}$ respectively. The first pair of equations (\ref{eq25}) is similar to the equations  for the vortex dynamo \cite{21s}, \cite{24s}. Coefficients  $\alpha_{1}$, $\alpha _{2}$ serve to establish  a positive feedback between the velocity, which is the projection of the Coriolis force. The second pair of equations  (\ref{eq26}) is similar to the well-known \cite{1s}-\cite{10s} dynamo theory for the  $\alpha$ - effect, which describes the enhancement of a large-scale magnetic field by small-scale spiral turbulence. In the system (\ref{eq26}) by means of the coefficients $\alpha_{H}^{(1)}$, $\alpha _{H}^{(2)}$ a positive feedback is also established for the component of the magnetic field that is due to the projections of the Coriolis force.
To study the large-scale instability described by equation system (\ref{eq25})-(\ref{eq26}), we choose perturbations in the form of plane waves with wave vector $\vec{K} \parallel OZ$, i.e.
\begin{equation} \label{eq29}
 \left(\begin{array}{c} {W_{1,2} } \\ {H_{1,2}
} \end{array}\right)=\left(\begin{array}{c} {A_{W_{1,2} } } \\ {A_{H_{1,2} } } \end{array}
\right){exp}\left(-i\omega T+iKZ\right)
\end{equation}
Substituting (\ref{eq29}) into equation system (\ref{eq25}), (\ref{eq26}),  we obtain the dispersion equation:

\begin{figure}
  \centering
  \includegraphics[height= 7 cm]{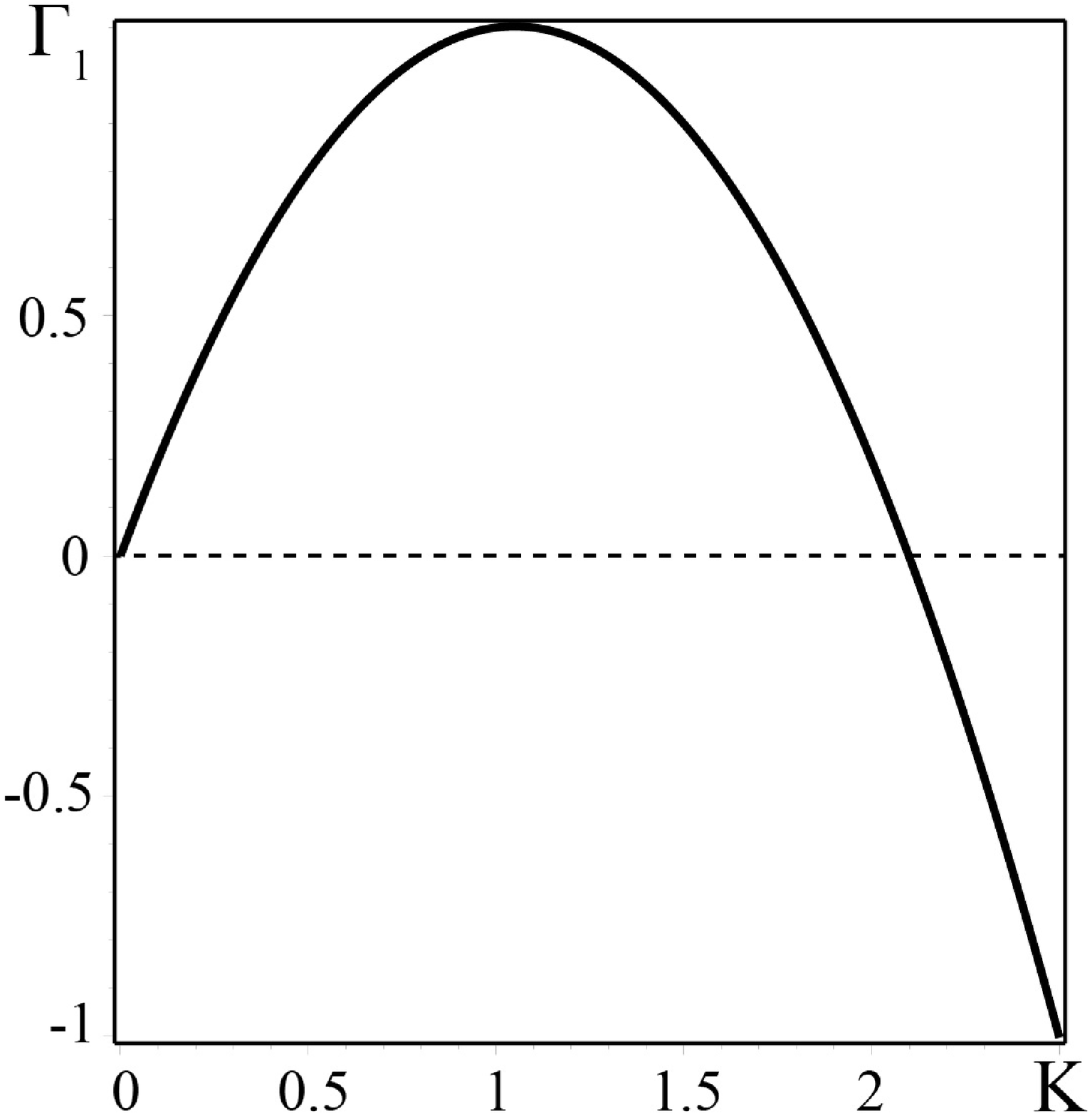}
	\includegraphics[height= 7 cm]{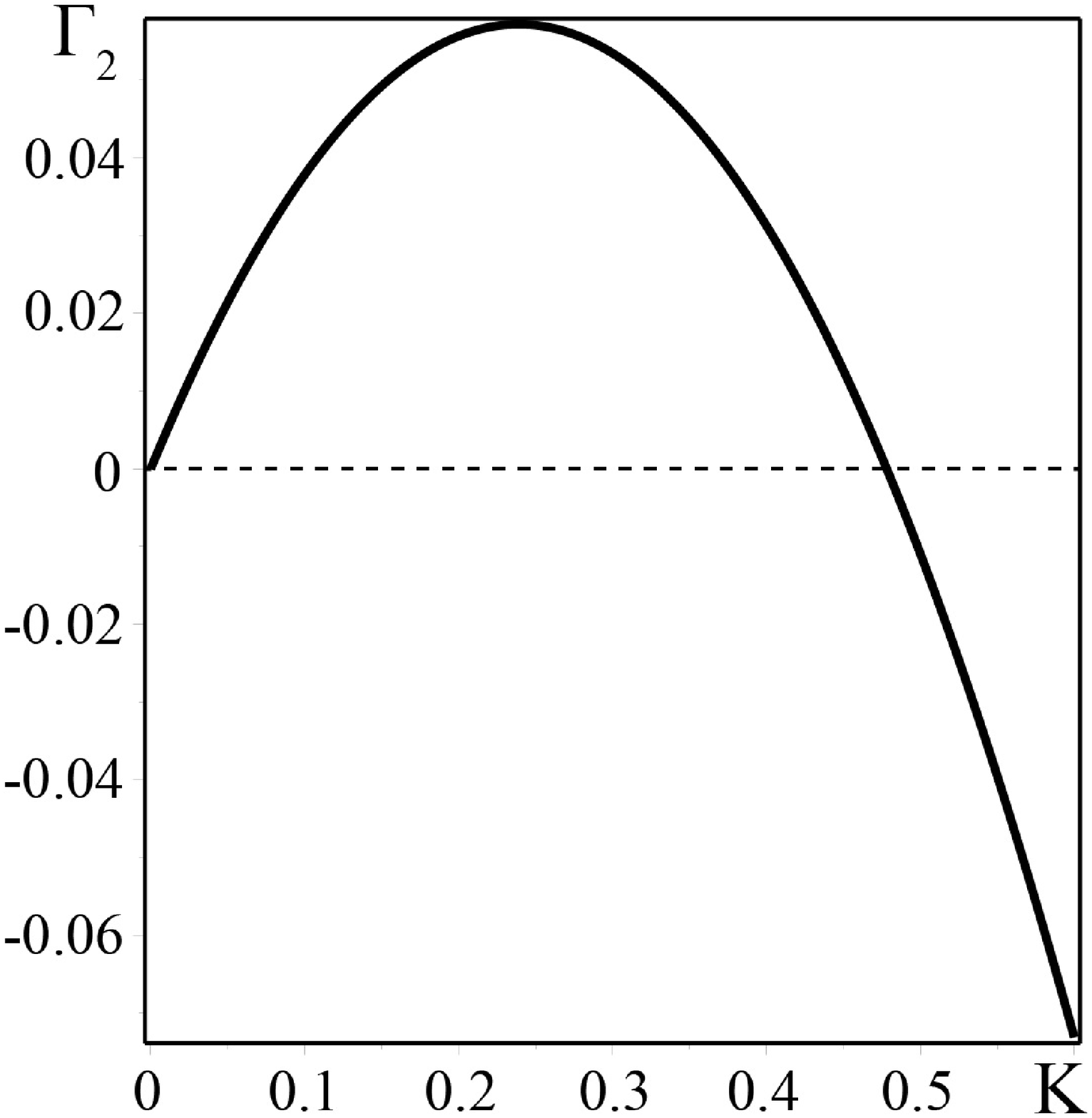}\\
\caption{On the left is a plot of the dependence of the instability increment for  HD  $\alpha $--effect on the wave numbers $K$  for the rotation parameter $D=2.5$; on the right is a plot  of the dependence of the instability increment for the MHD $\alpha $--effect on the wave numbers   $K$ with the rotation parameter  $D=1.5$ and the magnetic Prandtl number  $Pm=1$.}\label{fg3}
\end{figure}

\begin{equation} \label{eq30}
 \left(-i\omega +K^{2} \right)^{2} -\alpha _{1}
\alpha _{2} K^{2} =0,  \left(-i\omega +Pm^{-1} K^{2} \right)^{2} -\alpha _{H}^{(1)}
\alpha _{H}^{(2)} K^{2} =0
\end{equation}
Substituting   $\omega =\omega _{0} +i \Gamma$ into equations (\ref{eq30}),   we find:
\begin{equation} \label{eq31}
 \Gamma_{1} =Im \omega _{1} =\pm \sqrt{
\alpha _{1} \alpha _{2} } K-K^{2}
\end{equation}

\begin{equation} \label{eq32}
 \Gamma_{2} =Im\omega _{2} =\pm \sqrt{
\alpha _{H}^{(1)} \alpha _{H}^{(2)} } K-Pm^{-1} K^{2}
\end{equation}
The solution (\ref{eq31}) points to the existence of an instability at   ${\alpha _{1} \alpha _{2} } >0$ for large-scale eddy perturbations with a maximum instability increment of $\Gamma_{1max} =\frac{\alpha _{1} \alpha _{2}}{4}$ at wave numbers $K_{1max} =\frac{\sqrt{\alpha _{1} \alpha _{2} } }{2}$. Similarly, for magnetic disturbances, the  growth rate of the instability  $\Gamma_{2max} =\frac{\alpha _{H}^{(1)} \alpha _{H}^{(2)}}{4} Pm$   reaches its maximum  value at wave numbers $K_{2max} =\frac{\sqrt{\alpha_{H}^{(1)} \alpha _{H}^{(2)} } }{2} Pm$. If   ${\alpha _{1} \alpha _{2} } <0$ and ${\alpha _{H}^{(1)} \alpha _{H}^{(2)}} <0$, then, instead of the instabilities, damped oscillations arise, with the frequencies $\omega _{01} =\sqrt{\alpha _{1} \alpha _{2} } K$ and $\omega_{02} =\sqrt{\alpha _{H}^{(1)} \alpha _{H}^{(2)}} K$ respectively. Let us note that in the linear theory considered here,  the coefficients $\alpha_{1}$, $\alpha _{2}$, $\alpha _{H}^{(1)}$, $\alpha_{H}^{(2)}$ do not depend on the amplitudes of the fields, but depend only on the rotation parameters $D_{1,2}$, the magnetic Prandtl number $Pm$, and the amplitude of the external force $f_{0}$ . Let us analyze the dependence of these coefficients on dimensionless parameters, assuming for simplicity the dimensionless amplitude of the external force $f_{0}$  equal to $ f_{0} =10$. Fixing the level of a dimensionless force means choosing a certain level of stationary background of small-scale and rapid oscillations.
In the expressions for  coefficients  $\alpha _{1}$, $\alpha _{2}$, $\alpha _{H}^{(1)}$, $\alpha _{H}^{(2)}$  instead of Cartesian projections  $D_{1,2}$   it is convenient to go over to their projections in the spherical coordinate system $(D,\phi,\theta)$. The coordinate surface  $D=const$ is a sphere, $\theta$ is the latitude: $\theta \in [0,\pi ]$ ,  $\phi$    is the longitude: $\phi \in [0,2\pi ]$ .  We analyze the dependence of the coefficients  $\alpha _{1}$, $\alpha _{2}$, $\alpha _{H}^{(1)}$, $\alpha _{H}^{(2)}$  from the rotation effect, assuming  for simplicity $D_{1}=D_{2}$, which corresponds to a fixed value of longitude  $\phi =\pi /4+\pi n$,  where $n=0,1,2...k$, $k$ is an integer number. In this case, the amplification coefficients of the vortex and magnetic disturbances are respectively
\[ \alpha =\alpha _{1} =\alpha _{2} =f_{0}^{2}\frac{8\sqrt{2}D\sin \theta ({{D}^{2}}{{\sin }^{2}}\theta -4)}{{{({{D}^{4}}{{\sin }^{4}}\theta +16)}^{2}}} ,\]

\[\alpha _{H} =\alpha
_{H}^{(1)} =\alpha _{H}^{(2)} =f_{0}^{2}\frac{2\sqrt{2}D\sin \theta Pm}{(1+P{{m}^{2}})({{D}^{4}}{{\sin }^{4}}\theta +16)}\]

It can be seen from these expressions, that, at  the poles $(\theta=0,\; \theta=\pi)$   the generation of vortex and magnetic disturbances is not effective because  $\alpha ,{{\alpha }_{H}}\to 0$, i.e. large-scale instability arises if the angular velocity vector of rotation $\vec\Omega$ is deviated from  $Z$ axis. The dependence of the ratio $\alpha$ from the rotation of the fluid (parameter $D$) at a fixed latitude $\theta =\pi /2$ is depicted in the left part of Fig. \ref{fg2}. This shows that  ${\alpha}_{\max} $  achieves its most negative value  at  $D=1$. In this case, the rising of decaying mode occurs. Further, when  the parameter  $D$ is increasing, $\alpha $  shifts in positive direction, passing through zero at  $\alpha=0 $  for $D=2$. At further increase of  parameter $D $,  the coefficient $\alpha $, after reaching its maximum value  $\alpha_{\max}$,  smoothly tends to zero, i.e. fast rotation suppresses HD  $\alpha $ - effect. A similar phenomenon was described in \cite{43s}. Let us now consider the dependence of the coefficient $\alpha _{H} $ on the rotation parameter $D$, assuming latitude $\theta =\pi /2$ and  magnetic Prandtl number $Pm=1$.
MHD   $\alpha $ - effect (or $\alpha _{H} $ - effect)  is also increasing  at  slow  rotation to a maximum value of $\alpha_{Hmax} $ . After  that,  the increase of  the parameter $D$, is accompanied  by decline of $\alpha _{H} $, but the sign of the coefficient  $\alpha _{H} $ does not change. Analysis of the dependence $\alpha _{H}(D) $ showed that fast environment rotation also  suppresses MHD  $\alpha $ - effect (see right part of Fig. \ref{fg2}). Let us fixin the values of the rotation parameters $D$ and the magnetic Prandtl number  $Pm$   for wide angles $\theta =\pi /2$, and build  the plots of the dependence of the growth rate on the vortex  $\Gamma_{1}$ and magnetic $\Gamma_{2}$ perturbation on  the wave number $K$. These shape of these plots is typical for $\alpha $ - effect (see Fig. \ref{fg3}).
Thus, as a result of the development of  instability in an obliquely rotating electroconductive fluid, there are are generated  large-scale spiral circularly polarized vortices and magnetic fields  of Beltrami type.
\begin{figure}
\centering
    \includegraphics[height=6 cm]{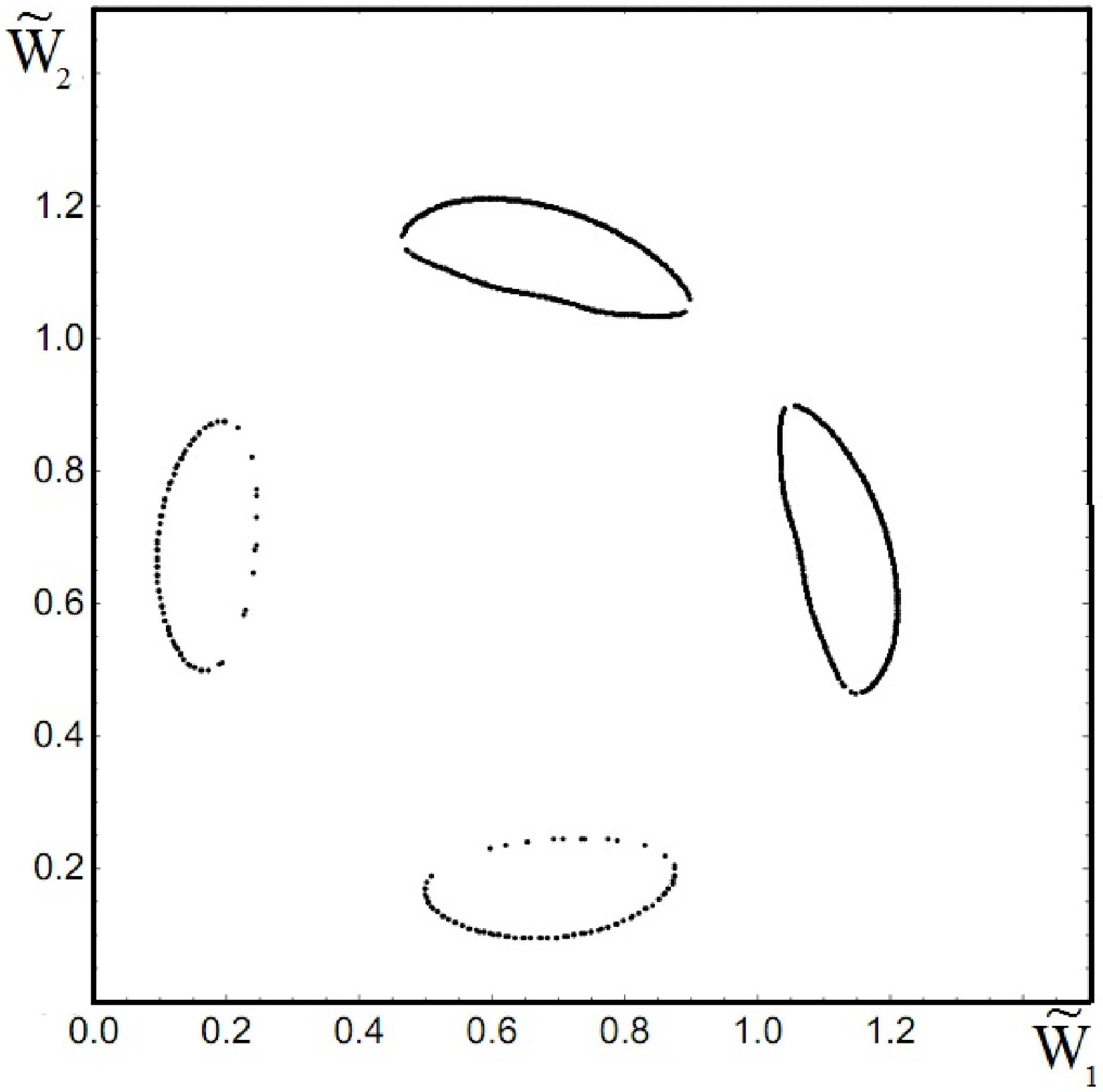}
		  \includegraphics[height=6 cm]{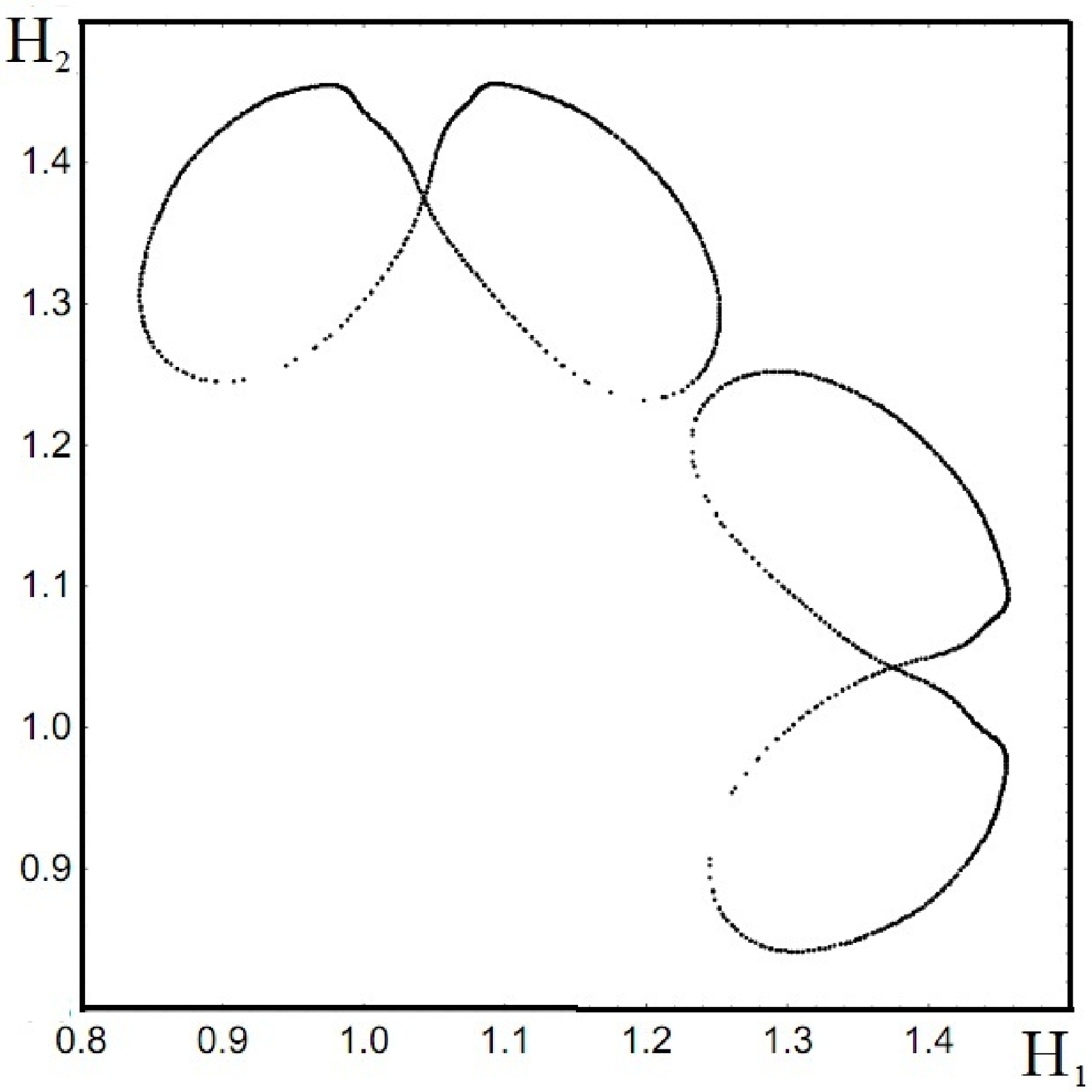}\\
	 \includegraphics[height=6 cm]{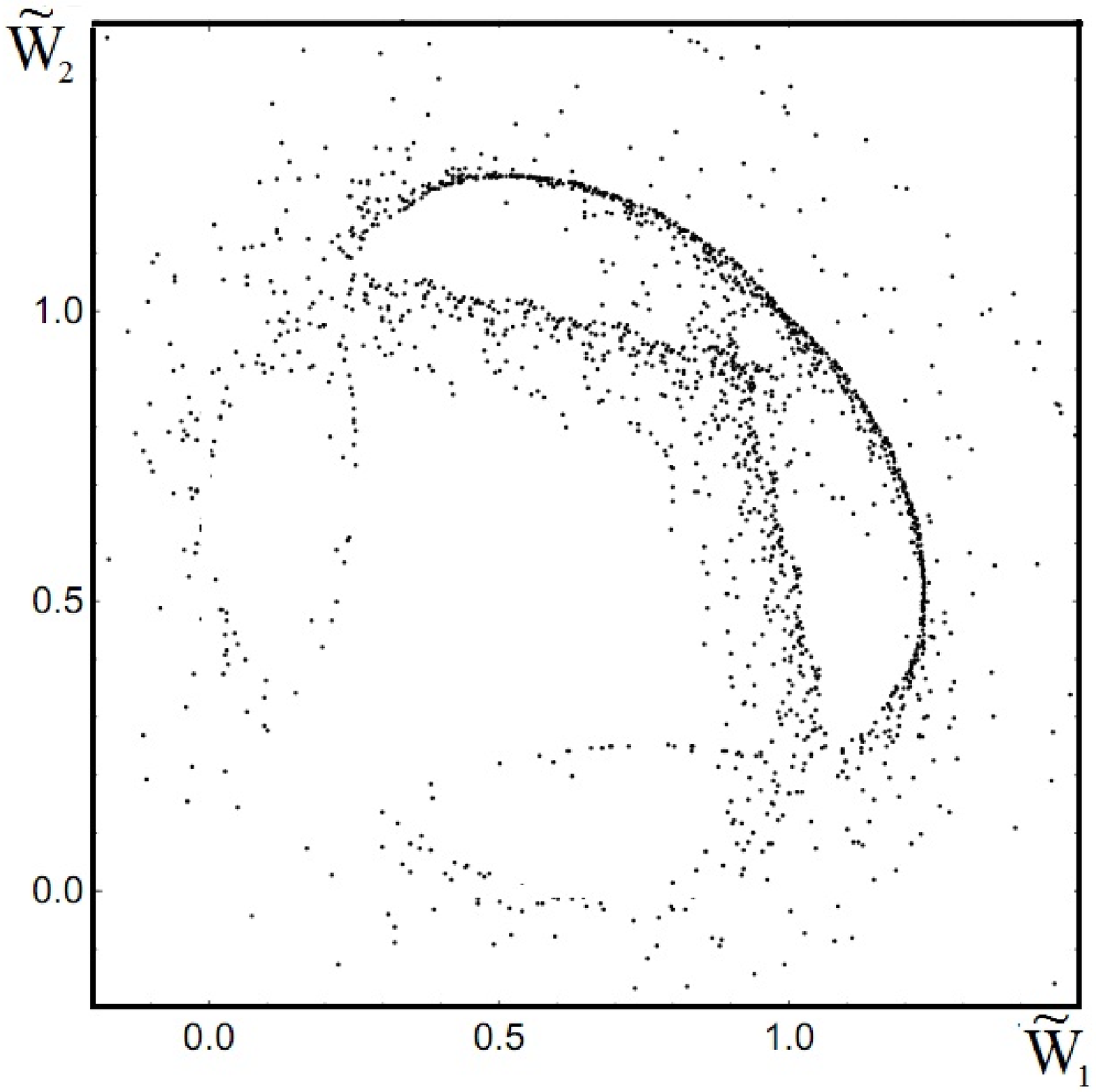}
		  \includegraphics[height=6 cm]{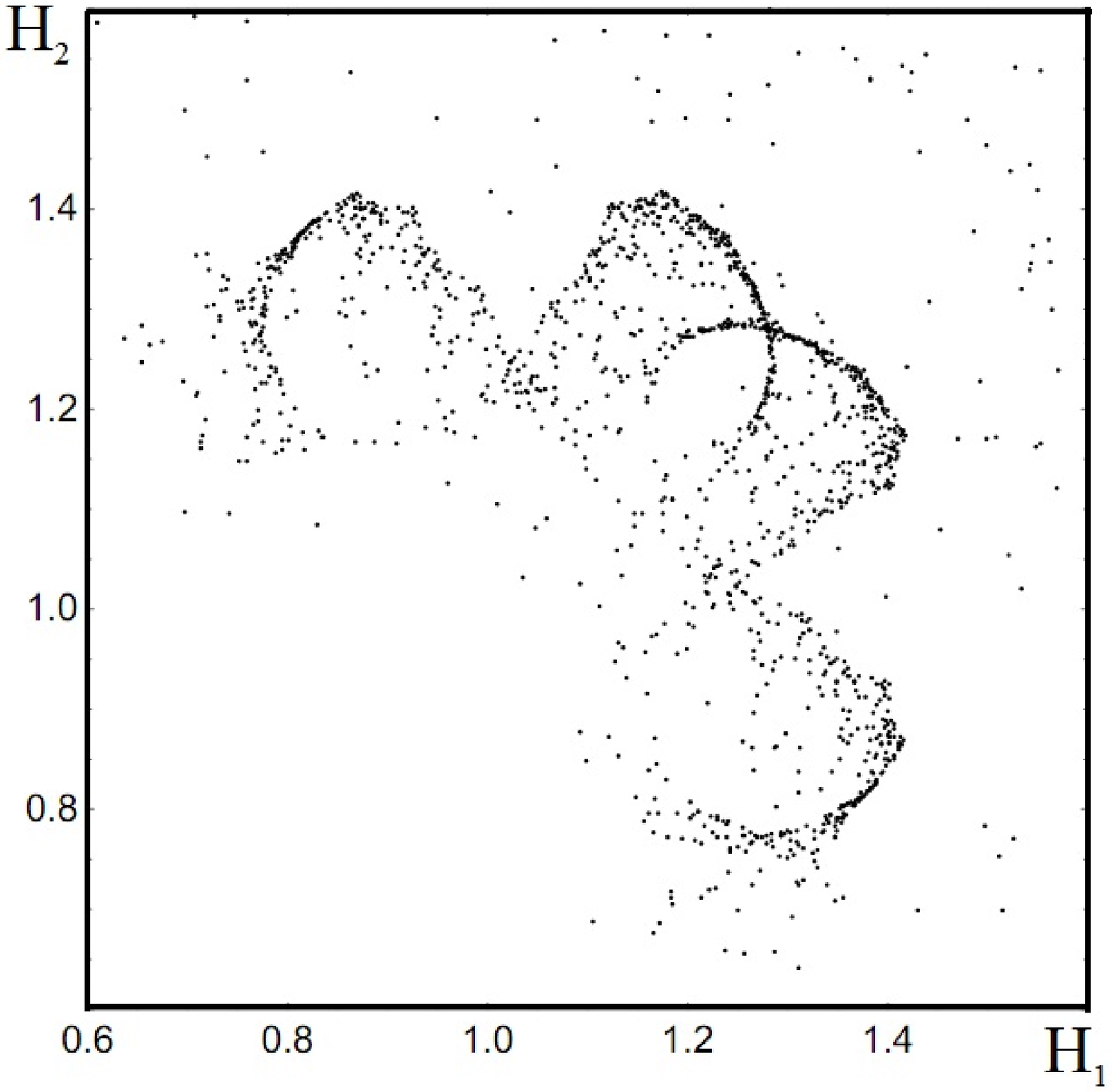}\\
  \caption{Poincare sections for  two phase trajectories. The upper one is  for the trajectory with initial conditions  $\widetilde W_1 (0) = 1$, $\widetilde W_2 (0) = 1$, $ H_1 (0) = 1.2$, $ H_2 (0) = 1.2$  while the lover one is for trajectory with initial conditions  $\widetilde W_1 (0) = 1$, $\widetilde W_2 (0) = 1$, $ H_1 (0) = 1.149$, $ H_2 (0) = 1.149$.  It is easy to see that the trajectory corresponding to the upper figures is wounded on  tori. This is a regular type of trajectory. The lower figures show stochastic layers, to which the corresponding chaotic trajectory belongs.}\label{fg4}
\end{figure}

\section{Nonlinear stationary structures}

It is obvious that with an increase in the perturbation amplitude  $W_{1,2} $ and $H_{1,2} $, the nonlinear coefficients  $\alpha _{\left(1\right)} $, $\alpha _{\left(2\right)} $, $\alpha _{H}^{\left(1\right)} $, $\alpha _{H}^{\left(2\right)} $  decrease  and  the instability saturates, passing to the stationary regime. As a result, nonlinear stationary structures are formed. To describe such structures, we consider the nonlinear system of equations \eqref{eq21}-\eqref{eq24} in the stationary case, assuming  ${\partial}_TW_1={\partial}_TW_2={\partial}_TH_1={\partial}_TH_2=0$  and integrating these equations in $Z$:
\begin{figure}
\centering
    \includegraphics[height=7 cm]{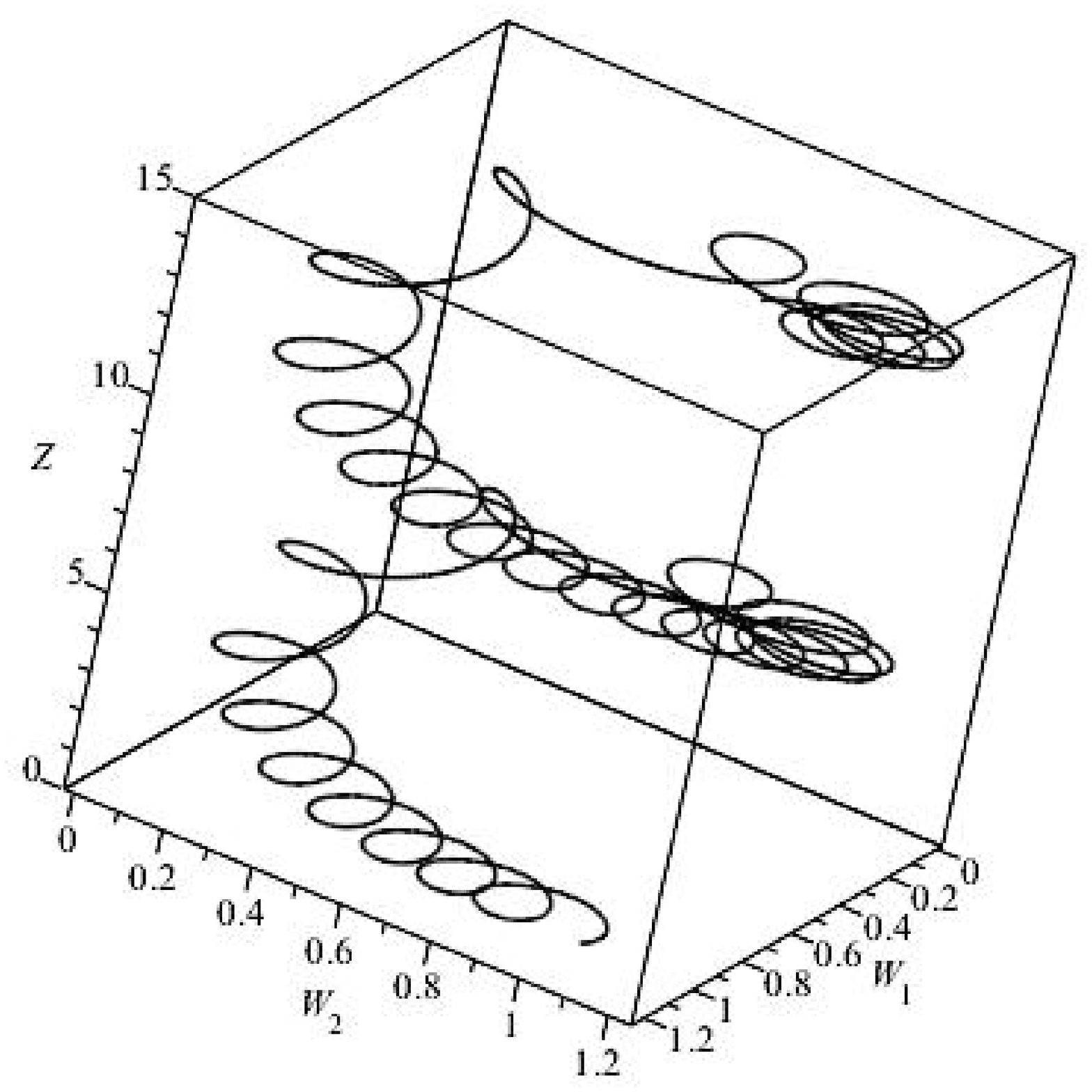}
		  \includegraphics[height=7 cm]{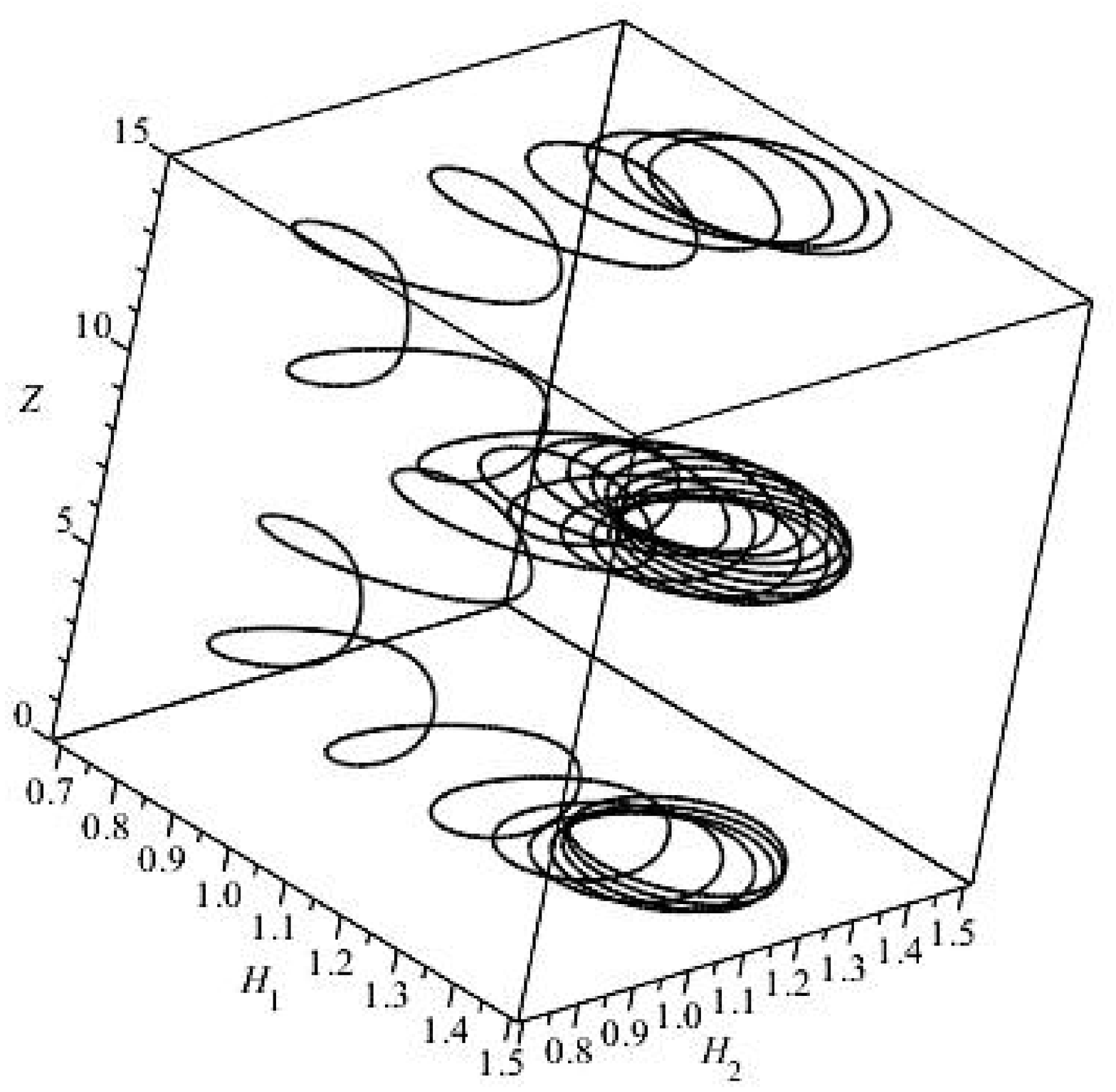}\\
    \includegraphics[height=7 cm]{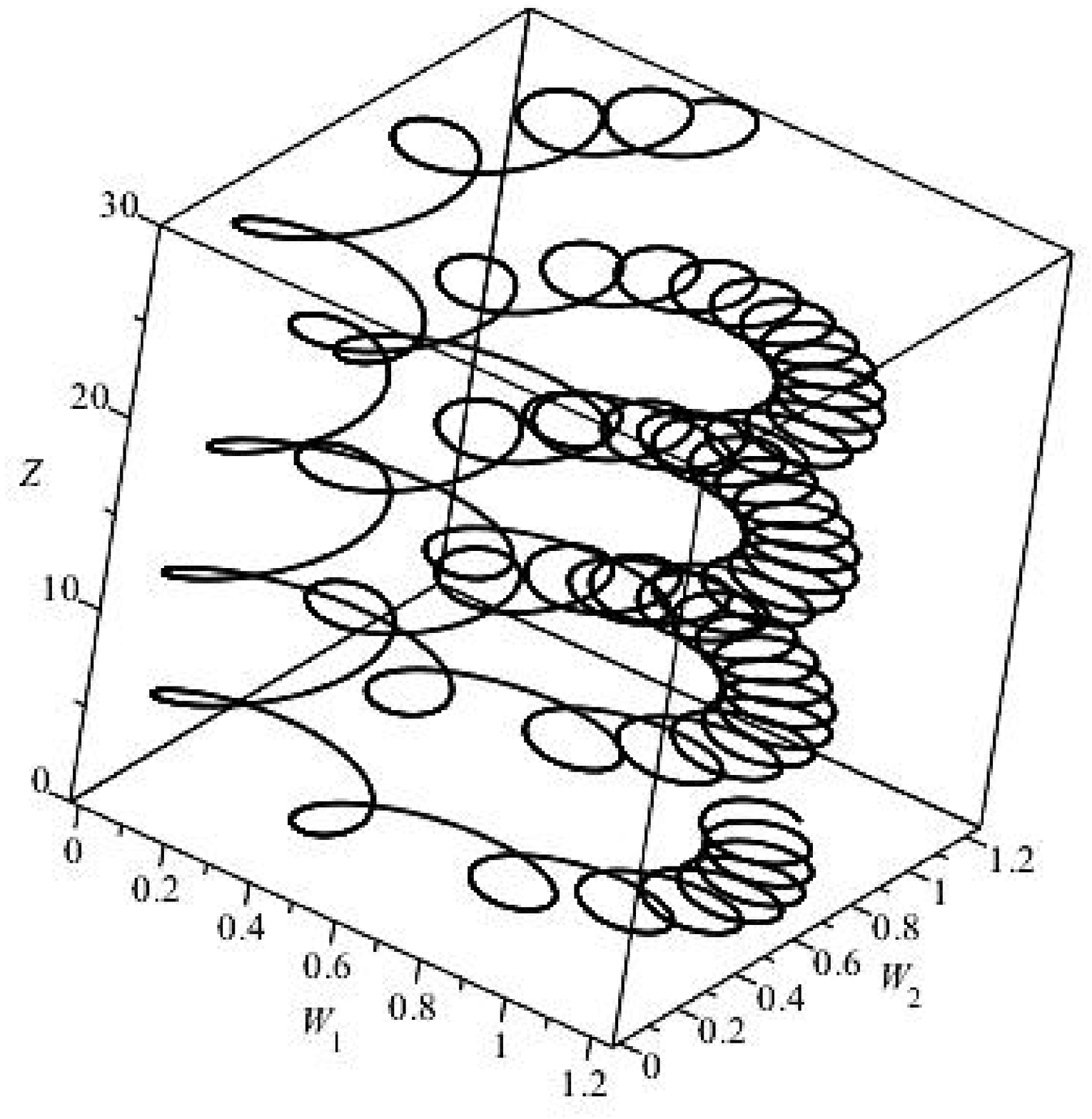}
  \includegraphics[height=7 cm]{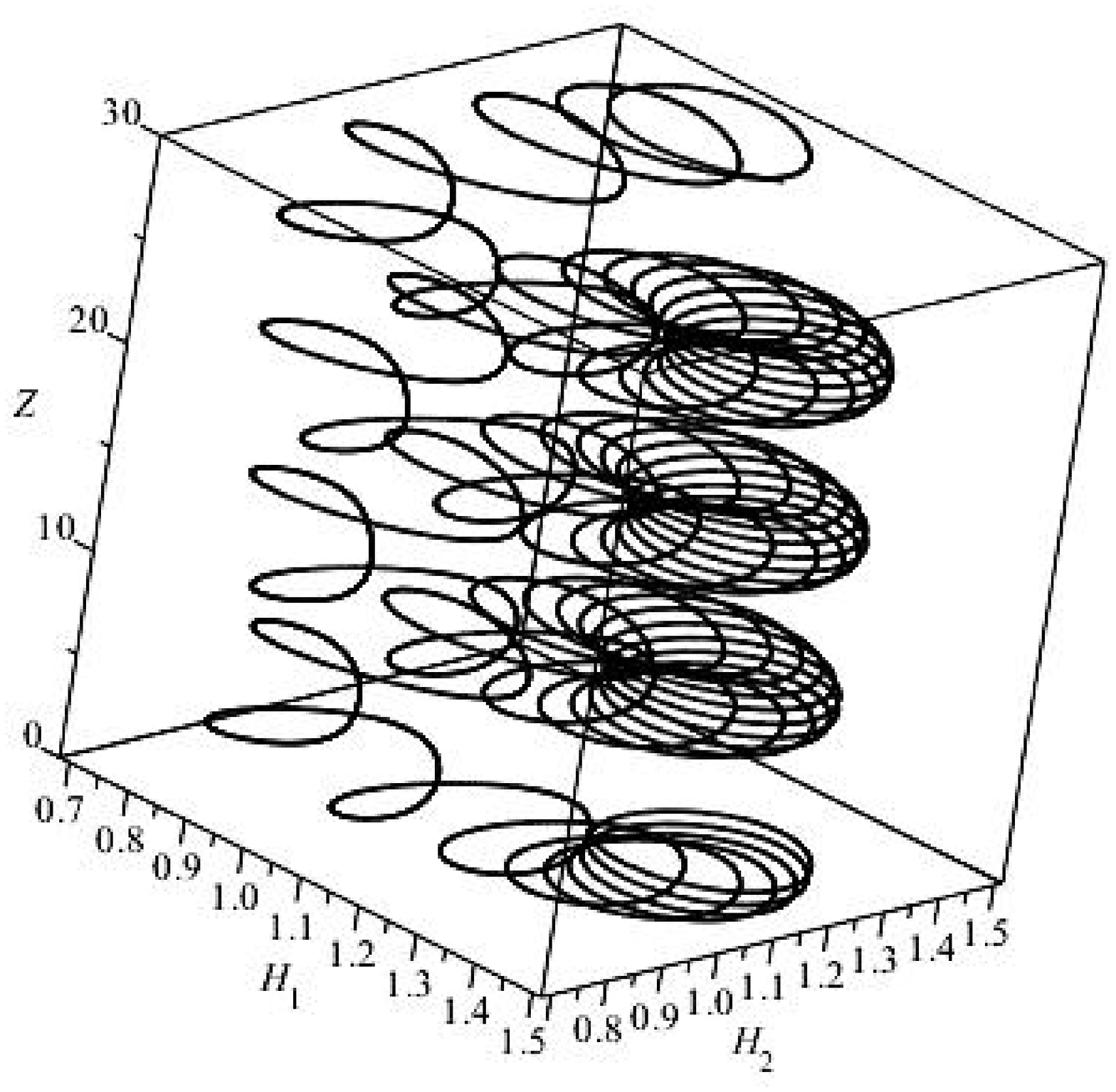}\\
  \caption{The upper part shows  the dependence of the velocity and magnetic field on the height for the numerical solution of equation system (\ref{eq33})-(\ref{eq36}) with the initial conditions: $\widetilde W_1 (0) = 1$, $\widetilde W_2 (0) = 1$, $ H_1 (0) = 1.2$, $ H_2 (0) = 1.2$. This dependence corresponds to regular motions of the Poincare section shown on top of Fig. \ref{fg4}.  Below a similar  dependence is shown  for the numerical solution of equation system (\ref{eq33})-(\ref{eq36}) with the initial conditions: $\widetilde W_1 (0) = 1$, $\widetilde W_2 (0) = 1$, $ H_1 (0) = 1.149$, $ H_2 (0) = 1.149$. This chaotic dependence corresponds to the Poincare sections in Fig. \ref{fg4} shown at the bottom.} \label{fg5}
\end{figure}

\begin{equation} \label{eq33}
  \frac{d{\widetilde{W}}_1}{dZ}=-{\alpha }_{\left(2\right)}{\widetilde{W}}_2+C_1
\end{equation}

\begin{equation} \label{eq34}
    \frac{d{\widetilde{W}}_2}{dZ}={\alpha }_{\left(1\right)}{\widetilde{W}}_1+C_2
\end{equation}

\begin{equation} \label{eq35}
    \frac{1}{Pm}\frac{dH_1}{dZ}=\alpha_{H}^{\left(2\right)}H_2+C_3
\end{equation}

\begin{equation} \label{eq36}
    \frac{1}{Pm}\frac{dH_2}{dZ}=-\alpha_{H}^{\left(1\right)}H_1+C_4
\end{equation}
Here we use the notation ${\widetilde{W}}_1=1-W_1$, ${\widetilde{W}}_2=1-W_2$; $C_1$, $C_2$, $C_3$ and $C_4$  are arbitrary integration constants. Like the previous section,  for the coefficients ${\alpha }_{\left(1\right)}$, ${\alpha }_{\left(2\right)}$, $\alpha_{H}^{\left(1\right)}$, $\alpha_{H}^{\left(2\right)}$  instead of  Cartesian projections  $D_{1,2}$   we use   projections in  the spherical coordinate system. Considering, for the sake of simplicity, fixed values of longitude  $\phi =\pi /4$   and latitude $\theta =\pi /2$, one obtains  the coefficients  ${\alpha }_{\left(1\right)}$, ${\alpha }_{\left(2\right)}$, $\alpha_{H}^{\left(1\right)}$, $\alpha_{H}^{\left(2\right)}$ in the following form:

\begin{figure}
\centering
 \includegraphics[height=8 cm, width=12 cm]{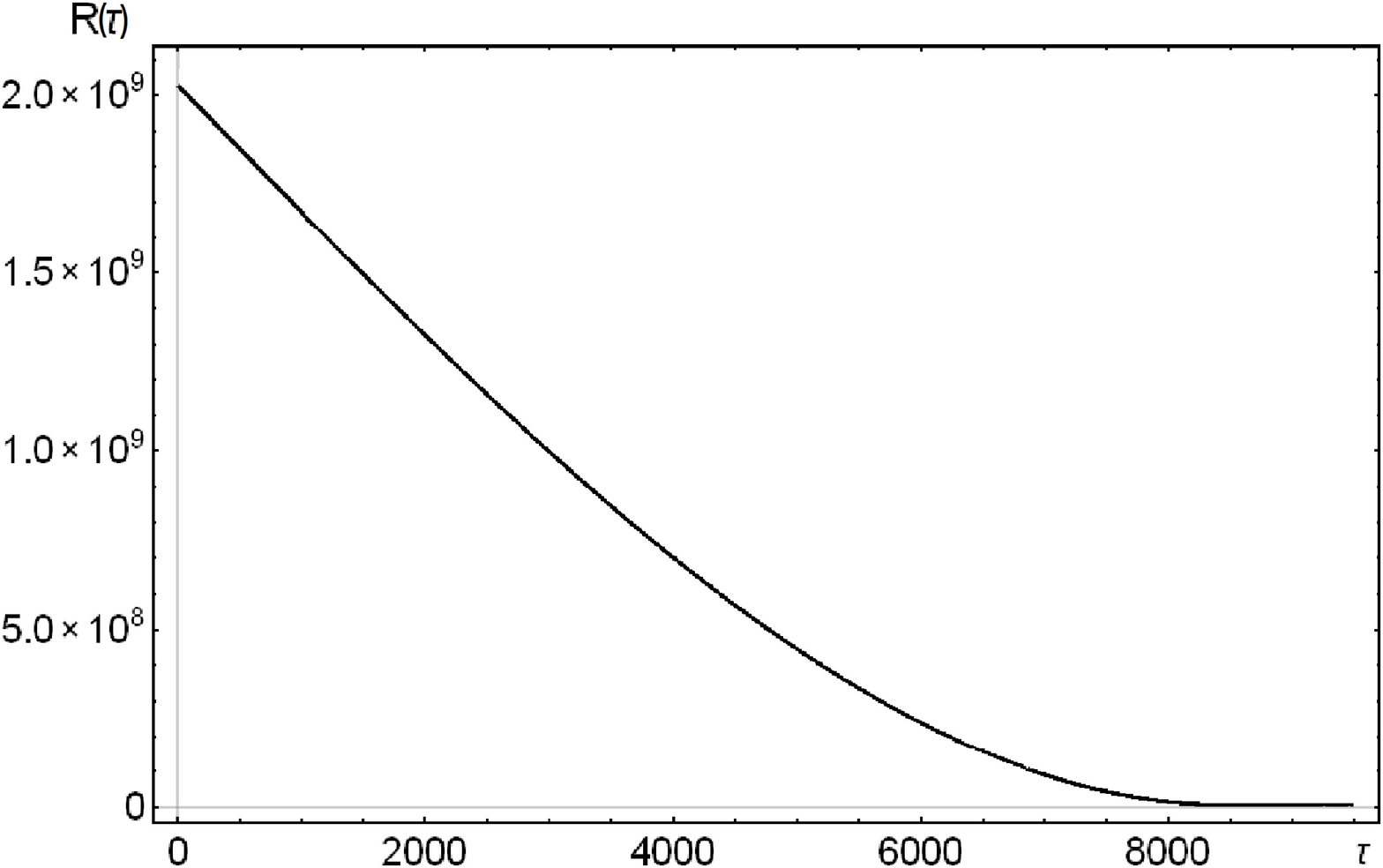}\\
	
  \caption{The plot of the dependence of the autocorrelation function  ${R}_{\widetilde W_1 \widetilde W_1}$   on time  $\tau$    for a trajectory with initial conditions  $\widetilde W_1 (0) = 1$, $\widetilde W_2 (0) = 1$, $ H_1 (0) = 1.149$, $ H_2 (0) = 1.149$  (chaotic motion).}\label{fg6}
\end{figure}

\[\alpha _{\left(1\right)} ={f_{0}^{2} } \frac{\sqrt{2} D q_{1} Q_{1}{{\widetilde W}_1}^{-1}  }{16
{{\widetilde W}_1}^2 q_{1}^{2} Q_{1}^{2} +\left[D^{2} + 2\left(1-{{\widetilde W}_1}^2
\right)+2\mu_{1} \right]^{2} } \]
\[\alpha _{\left(2\right)} ={f_{0}^{2} } \frac{\sqrt{2} D q_{2} Q_{2}{{\widetilde W}_2}^{-1}  }{16
{{\widetilde W}_2}^2 q_{2}^{2} Q_{2}^{2} +\left[D^{2} + 2\left(1-{{\widetilde W}_2}^2
\right)+2\mu_{2} \right]^{2} } \]
\[\alpha _{H}^{\left(1\right)}={f_{0}^{2}} \frac{4 \sqrt{2} D Pm {\widetilde W}_1 Q _{1} }{\left(1+Pm^{2}{{\widetilde W}_1}^{2}\right)\left[16{{\widetilde W}_1}^{2} q_{1}^{2} Q_{1}^{2} +\left[D^{2} +2 \left(1-{{\widetilde W}_1}^2 \right)+2\mu
_{1} \right]^{2} \right]}\]
\[\alpha _{H}^{\left(2\right)}={f_{0}^{2}} \frac{4 \sqrt{2} D Pm {\widetilde W}_2 Q _{2} }{\left(1+Pm^{2}{{\widetilde W}_2}^{2}\right)\left[16{{\widetilde W}_2}^{2} q_{2}^{2} Q_{2}^{2} +\left[D^{2} +2 \left(1-{{\widetilde W}_2}^2 \right)+2\mu
_{2} \right]^{2} \right]}\]
Equations (\ref{eq33})-(\ref{eq36}) constitute a nonlinear dynamical system in four-dimensional phase space. It is not difficult to see that this equation system is conservative.  However, the search for the Hamiltonian of this nonlinear system is technically a cumbersome task. The integration is complicated by the dependence of the nonlinear coefficients ${\alpha }_{\left(1\right)}$, ${\alpha }_{\left(2\right)}$, $\alpha_{H}^{\left(1\right)}$, $\alpha_{H}^{\left(2\right)}$ on the fields $\vec{W}$, $\vec{H}$  that takes it beyond the class of elementary functions. Full qualitative analysis of this equation system is exceptionally complicated due to the high dimension of the phase space, and a large number of parameters involved. On the basis of common concepts  of  such  systems of conservative equations, we should expect the presence of the structure of resonant and non-resonant tori in the phase space, and. as a consequence, the existence of a stationary chaotic structures of hydrodynamic and magnetic fields. One way to study such a complex system of nonlinear equations (\ref{eq33})-(\ref{eq36})  is a method of construction of Poincare sections. Choosing dimensionless parameters  $D=1.5$, $Q=Pm=1$,  $f_{0}=10$    and constants  ${C}_{1}=1, {C}_{2}=-1$, $ {C}_{3}=-1, {C}_{4}=1$ it is possible to construct numerically, using standard Mathematica programs,  Poincare sections of trajectories in the phase space.

\begin{figure}
\centering
 \includegraphics[height=8 cm, width=12 cm]{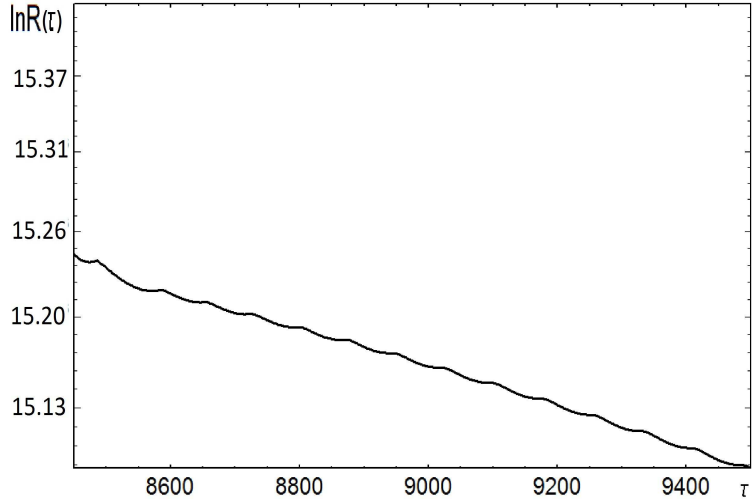}\\
	
  \caption{A rectilinear dependence of the autocorrelation function  ${R}_{\widetilde W_1 \widetilde W_1}$  in logarithmic scales  from the time interval $\tau$ for strongly chaotic motion.}\label{fg7}
\end{figure}
The upper part of Fig. \ref{fg4} demonstrates the Poincare section of a regular trajectory for the velocity field and the magnetic field. At the lower section, the structure of the chaotic layer, to which the chosen trajectory just  belongs, is clearly visible,. The presence of such chaotic trajectories  signifies  the existence of stationary chaotic structures, such  as velocity field and a magnetic field. In Fig. \ref{fg5}  the dependence is shown of stationary large-scale fields on the height, which was obtained numerically for the initial conditions corresponding  to Poincare sections in Fig. \ref{fg4}. From these figures we can also see the arising  of stationary chaotic solutions for the magnetic and vortex fields. In the numerical solution of equation system (\ref{eq33})-(\ref{eq36}),  the arising of  chaotic structures, was observed with a decrease of  the amplitudes of the initial field  and of the magnetic field. In addition to the Poincare section method, we use the notion of an autocorrelation function to prove the arising of a chaotic regime of stationary large-scale fields. As it is known (see, for example, \cite{44s}), the autocorrelation function $R(\tau)$  is used as a quantity characterizing the intensity of chaos, and, by definition, is the product of  random functions  $P(t)$ ,   at time moment $t$  and  $P(t+\tau)$    at  time moment  $t+\tau$,  averaged over the large time interval $\Delta t$:  $R(\tau ) = \mathop {\lim }\limits_{\Delta t \to \infty } \frac{1}{{\Delta t}}\int\limits_0^{\Delta t} {P(t)P(t+\tau )dt} $.
In the present work, the role of time  $t$  is played by the coordinate  $Z$ , and the product  $P(t)P(t + \tau )$  consists of sixteen components:
\[
P(t)P(t + \tau ) = \left[ {\begin{array}{*{20}c}
   {\widetilde W_1 (t)}  \\
   {\widetilde W_2 (t)}  \\
   {H_1 (t)}  \\
   {H_2 (t)}  \\
\end{array}} \right]\left[ {\begin{array}{*{20}c}
   {\widetilde W_1 (t + \tau )} & {\widetilde W_2 (t + \tau )} & {H_1 (t + \tau )} & {H_2 (t + \tau )}  \\
\end{array}} \right]
\]
Using known programs in Mathematica,  the  plot is obtained for the dependences  of autocorrelation functions for components of ${R}_{\widetilde W_1 \widetilde W_1}$ from time $\tau$, as shown in Fig. \ref{fg6}. The case of chaotic motion corresponds,  in Fig. \ref{fg6}, to the trajectory with exponential decay  functions  ${R}_{\widetilde W_1 \widetilde W_1}$. Clearly, the area of exponential decay in logarithmic scale of the autocorrelation function of ${R}_{\widetilde W_1 \widetilde W_1}$ can be approximated by a straight line (see Fig. \ref{fg7}). The data, presented in Fig. \ref{fg7}, allow us to determine the characteristic correlation time  ${\tau}_{cor}\approx69$   of  a stationary random process ${P}_{\widetilde W_{1}}$. If we go over to the above definition of time  $t$,  it becomes clear that we found the estimated value of height  ${Z}_{cor}\approx 69$  from which the chaotic motion starts of the stationary large-scale fields. At the bottom of Fig. \ref{fg5}  the chaotic solutions are shown for the velocity fields and magnetic fields up to height $Z=40$, which is less than ${Z}_{cor}$. However, in this case, we can see the beginning of a difficult confusing path to large-scale fields with increasing height $Z$. Visualization of such complex and confusing trajectories is very difficult.
Thus, with the increase of altitude  $Z$ up to some critical value ${Z}_{cor}$, quasi-periodic motion of large-scale stationary field is replaced by a chaotic.

\section{ Conclusion }

In this paper we obtained a closed system of nonlinear equations for vortex and magnetic large-scale perturbations (magneto-vortex dynamo) in an obliquely rotating electroconductive liquid with an external non-spiral force. At the initial stage of the evolution of small perturbations, the equations of the magneto-vortex dynamo are decoupled into two subsystems: vortex and magnetic. Moreover, in the linear stage, the generation of large-scale vortex perturbations occurs due to the development of an instability of the hydrodynamic type of  $\alpha $ - effect, and the generation of large-scale magnetic fields due to the instability of the MHD type of   $\alpha $ - effect. Both instabilities arise as a result of the combined action of the small-scale external force and the Coriolis force. A necessary condition for the appearance of instabilities here is the deviation of the angular velocity vector of rotation $\vec \Omega$ from the vertical axis $OZ$. In the general case, the coefficients of amplification  for HD and MHD $\alpha $ - effects depend on the magnitude of the angles -- $\phi$ longitude and  latitude $\theta$.  The minimum values of $\alpha $  correspond to the latitudes  $\theta \rightarrow 0$ ,  $\theta \rightarrow \pi$   (near the poles), and the maximum for $\theta \rightarrow \pi/2$ (near the equator).  Analysis of the effect of rotation on the growth vortex and magnetic disturbances showed,  that fast  rotation of the medium leads to damping of large-scale disturbances.
As the amplitude of the vortex and magnetic perturbations increases, the instability stabilizes and becomes stationary. In this regime, nonlinear stationary vortex and magnetic structures are formed. The dynamical system of equations describing these structures is Hamiltonian in the four-dimensional phase space. Numerical methods have proved the possibility of the existence of large-scale chaotic vortex and magnetic fields in the stationary regime.

\section*{Appendix I. Multiscale asymptotic developments}

Let us find the algebraic structure of the asymptotic development of the equations (\ref{eq7})-(\ref{eq9})   up to various orders of  $R$ , starting with the lowest one.   In order $R^{-3}$ there is only one equation
\begin{equation}\label{eqI1}
    \partial _{i} P_{-3} =0 \quad \Rightarrow \quad P_{-3}=P_{-3} \left( X \right)
\end{equation}
In order $R^{-2}$, equation arises:
\begin{equation}\label{eqI2}
    \partial _{i} P_{-2} =0 \quad \Rightarrow \quad P_{-2} =P_{-2} \left(X
\right)
\end{equation}
The equations (\ref{eqI1}) and (\ref{eqI2}) are satisfied automatically because   $P_{-3} $  and $P_{-2}$ are functions of only slow variables. In order $R^{-1}$ we obtain the system of equations
\[\partial _{t} W_{-1}^{i} +W_{-1}^{k} \partial _{k} W_{-1}^{i} =-\partial _{i} P_{-1}
-\nabla _{i} P_{-3} +\partial _{k}^{2} W_{-1}^{i} +\varepsilon _{ijk} W_{j} D_{k}
+\widetilde{Q}\varepsilon _{ijk} \varepsilon _{jml} \partial _{m} B_{-1}^{l} B_{-1}^{k} \]

\[ \partial_{t} B_{-1}^{i} -Pm^{-1} \partial _{k}^{2} B_{-1}^{i} =\varepsilon _{ijk} \varepsilon
_{knp} \partial _{j} W_{-1}^{n} B_{-1}^{p}  \]

\begin{equation}\label{eqI3}
    \partial
_{i} W_{-1}^{i} =0, \qquad \partial _{i} B_{-1}^{i}=0
\end{equation}
The averaging of equations (\ref{eqI3}) over on the <<fast>> variables give the secular equation
\begin{equation}\label{eqI4}
    -\nabla _{i} P_{-3} +\varepsilon _{ijk} W_{j} D_{k} =0
\end{equation}
which corresponds to geostrophic equilibrium.
In zero order in $R$ we have the equations:
\[\partial _{t} v_{0}^{i} +W_{-1}^{k} \partial _{k} v_{0}^{i} +v_{0}^{k} \partial
_{k} W_{-1}^{i} =-\partial _{i} P_{0} -\nabla _{i} P_{-2} +\partial _{k}^{2} v_{0}^{i}
+\varepsilon _{ijk} v_{0}^{j} D_{k} + \]

\[+ \widetilde{Q}\varepsilon _{ijk} \varepsilon
_{jml} \left(\partial _{m} B_{-1}^{l} B_{0}^{k} +\partial _{m} B_{0}^{l} B_{-1}^{k}
\right)+F_{0}^{i}\]

\[\partial _{t} B_{0}^{i} -Pm^{-1} \partial _{k}^{2} B_{0}^{i} =\varepsilon _{ijk}
\varepsilon _{knp} \; \left(\partial _{j} W_{-1}^{n} B_{0}^{p} +\partial _{j} v_{0}^{n}
B_{-1}^{p} \right)\]

\begin{equation}\label{eqI5}
    \partial _{i} v_{0}^{i} =0, \quad
\partial _{i} B_{0}^{i} =0
\end{equation}
These equations give one secular equation:
\[\nabla P_{-2} =0 \quad \Rightarrow \quad P_{-2} =const\]
Let us consider the  first order approximation $R^{1} $ :
\[\partial _{t} v_{1}^{i} +W_{-1}^{k} \partial _{k} v_{1}^{i} + v_{0}^{k}
\partial _{k} v_{0}^{i} + v_{1}^{k} \partial_{k} W_{-1}^{i} +W_{-1}^{k} \nabla _{k}
W_{-1}^{i} =-\nabla _{i} P_{-1} -\]

\[- \partial _{i} \left(P_{1} +\overline{P}_{1} \right)+
\partial _{k}^{2} v_{1}^{i} + 2\partial _{k} \nabla _{k}
W_{-1}^{i} +{\varepsilon}_{ijk} {v}_{1}^{j} {D}_{k}+
\]

\begin{equation}\label{eqI6}
    +\widetilde{Q}\varepsilon _{ijk} \varepsilon _{jml} \left(\partial _{m} B_{-1}^{l} B_{1}^{k}
+\partial _{m} B_{0}^{l} B_{0}^{k} +\partial _{m} B_{1}^{l} B_{(-1)}^{k} +\nabla
_{m} B_{(-1)}^{l} B_{(-1)}^{k} \right)
\end{equation}

\[\partial _{t} B_{1}^{i} -Pm^{-1} \partial _{k}^{2} B_{1}^{i} -Pm^{-1} 2\partial
_{k} \nabla _{k} B_{-1}^{i} =\varepsilon _{ijk} \varepsilon _{knp} \left(\partial _{j}
W_{-1}^{n} B_{1}^{p} +\partial _{j} v_{0}^{n} B_{0}^{p} +\partial _{j} v_{1}^{n}
B_{-1}^{p} +\nabla _{j} {W}_{-1}^{n} {B}_{-1}^{p} \right)\]

\[\partial
_{i} v_{1}^{i} +\nabla _{i} W_{-1}^{i} =0, \quad \partial
_{i} B_{1}^{i} +\nabla _{i} B_{-1}^{i} =0\]
From this system of equations,  secular equations follow:
\begin{equation}\label{eqI7}
    W_{-1}^{k} \nabla _{k} W_{-1}^{i} =-\nabla _{i} P_{-1} +\widetilde{Q}\varepsilon
_{ijk} \varepsilon _{jml} \nabla _{m} B_{-1}^{l} B_{-1}^{k}
\end{equation}
\begin{equation}\label{eqI8}
    \varepsilon
_{ijk} \varepsilon _{knp} \nabla _{j} W_{-1}^{n} B_{-1}^{p} =0
\end{equation}

\begin{equation}\label{eqI9}
    \nabla_{i} W_{-1}^{i} =0 \quad \nabla _{i} B_{-1}^{i} =0
\end{equation}
In the second order $R^{2}$ we obtain the equations:
\[\partial _{t} v_{2}^{i} +W_{-1}^{k} \partial _{k} v_{2}^{i} +v_{0}^{k}
\partial _{k} v_{1}^{i} +W_{-1}^{k} \nabla _{k} v_{0}^{i} +v_{0}^{k} \nabla _{k}
W_{-1}^{i} +v_{1}^{k} \partial _{k} v_{0}^{i} +v_{2}^{k} \partial _{k} W_{-1}^{i}
=\]
\[= -\nabla _{i} P_{2} -\nabla _{i} P_{0} +\partial _{k}^{2} v_{2}^{i} +2\partial _{k}
\nabla _{k} v_{0}^{i} +{\varepsilon }_{ijk} {v}_{2}^{j}
{D}_{k} +\]
\[+\widetilde{Q}\varepsilon _{ijk} \varepsilon _{jml} \left(\partial
_{m} B_{-1}^{l} B_{2}^{k} +\partial _{m} B_{0}^{l} B_{1}^{k} +\partial _{m} B_{1}^{l}
B_{0}^{k} +\partial _{m} B_{2}^{l} B_{-1}^{k} +\nabla _{m} B_{-1}^{l} B_{0}^{k} +
\nabla _{m} B_{0}^{l} B_{-1}^{k} \right)\]

\[\partial _{t} B_{2}^{i} -Pm^{-1} \partial _{k}^{2} B_{2}^{i} -Pm^{-1}
2\partial _{k} \nabla _{k} B_{0}^{i} = \varepsilon _{ijk} \varepsilon _{knp}
\left(\partial _{j} W_{-1}^{n} B_{2}^{p} +\partial _{j} v_{0}^{n} B_{1}^{p} +\right.\]

\[\left.+\partial_{j}  v_{2}^{n}  B_{-1}^{p} +
\nabla _{j} W_{-1}^{n} B_{0}^{p} +\nabla _{j} v_{0}^{n}  B_{-1}^{p} \right)\]

\begin{equation}\label{eqI10}
    \partial _{i}  v_{2}^{i} +\nabla _{i} v_{0}^{i} =0, \quad \partial _{i} B_{2}^{i} +\nabla _{i}  B_{0}^{i} =0
\end{equation}
It is easy to see that there are no secular terms in this order. Let us consider now the most important order $R^{3} $. In this order we obtain the equations:
\[\partial _{t} v_{3}^{i} +\partial _{T} W_{-1}^{i} +W_{-1}^{k}
\partial _{k} v_{3}^{i} +v_{0}^{k} \partial _{k} v_{2}^{i} +W_{-1}^{k} \nabla _{k}
v_{1}^{i} +v_{0}^{k} \nabla _{k} v_{0}^{i} +v_{1}^{k} \partial _{k} v_{1}^{i} +v_{1}^{k}
\nabla _{k} W_{-1}^{i} + v_{2}^{k} \partial _{k} W_{-1}^{i} =\]

\[=-\partial _{i} P_{3}
- \nabla _{i} \left(P_{1} +\overline{P}_{1} \right)+\partial _{k}^{2} v_{3}^{i}
+2\partial _{k} \nabla _{k} v_{1}^{i} +\Delta W_{-1}^{i} +{\varepsilon}_{ijk} {v}_{3}^{j} {D}_{k} +\]

\[+\widetilde{Q}\varepsilon _{ijk}
\varepsilon _{jml} \left(\partial _{m} B_{-1}^{l} B_{3}^{k} +\partial _{m} B_{0}^{l}
B_{2}^{k} +\partial _{m} B_{1}^{l} B_{1}^{k} +\partial _{m} B_{2}^{l} B_{0}^{k} +
\nabla _{m} B_{-1}^{l} B_{1}^{k} +\nabla _{m} B_{0}^{l} B_{0}^{k} \right)\]

\[\partial _{t} B_{3}^{i} +\partial _{T} B_{-1}^{i} -Pm^{-1} \partial _{k}^{2} B_{3}^{i}
-Pm^{-1} 2\partial _{k} \nabla _{k} B_{1}^{i} -Pm^{-1} \Delta B_{-1}^{i} =\]

\[=
\varepsilon _{ijk} \varepsilon _{knp} \left(\partial _{j} W_{-1}^{n} B_{3}^{p} +
\partial _{j} v_{0}^{n} B_{2}^{p} +\partial _{j} v_{1}^{n} B_{1}^{p} +\partial _{j}
v_{2}^{n} B_{0}^{p} +\nabla _{j} W_{-1}^{n} B_{1}^{p} +\nabla _{j} v_{0}^{n} B_{0}^{p}
\right)\]

\begin{equation}\label{eqI11}
    \partial
_{i} v_{3}^{i} +\nabla _{i} v_{1}^{i} =0, \quad \partial _{i} B_{3}^{i} +\nabla
_{i} B_{1}^{i} =0
\end{equation}
After averaging this system of equations over fast variables, we obtain the main system of secular equations to describe the evolution of large-scale perturbations:
\begin{equation}\label{eqI12}
    \partial _{t} W_{-1}^{i} -\Delta W_{-1}^{i} +\nabla _{k} \left(\overline{v_{0}^{k}
v_{0}^{i} }\right)=-\nabla _{i} \overline{P}_{1} +\widetilde{Q}\varepsilon _{ijk}
\varepsilon _{jml} \nabla _{m} \left(\overline{B_{0}^{l} B_{0}^{k} }\right)
\end{equation}
\begin{equation}\label{eqI13}
    \partial
_{t} B_{-1}^{i} -Pm^{-1} \Delta B_{-1}^{i} =\varepsilon _{ijk} \varepsilon _{knp}
\nabla _{j} (\overline{v_{0}^{n} B_{0}^{p} })
\end{equation}

\section*{Appendix II. Small-scale fields}
In Appendix I, we  obtained  the equations for  the asymptotic expansion in the zero approximation, which can be written in the following form:
\begin{equation} \label{eqII1}
 \widehat{D}_{W} v_{0}^{i} =-\partial _{i} P_{0}
+\varepsilon _{ijk} v_{0}^{j} D_{k} +\widetilde {Q}H_{k} \left(\partial _{k} B_{0}^{i}
-\partial _{i} B_{0}^{k} \right)+F_{0}^{i}
\end{equation}

\begin{equation} \label{eqII2}
\widehat{D}_{H} B_{0}^{i} =\left(H_{p} \partial
_{p} \right)v_{0}^{i}
\end{equation}

\begin{equation} \label{eqII3}
 \partial _{i} v_{0}^{i} =\partial _{k} B_{0}^{k}
 =0
\end{equation}
where we  introduced a notation for operators:
\[\widehat{D}_{W} =\partial _{t} -\partial _{k}^{2} +W_{-1}^k \partial _{k} ,\;\widehat{D}_{_{H}
} =\partial _{t} -Pm^{-1} \partial _{k}^{2} +W_{-1}^k \partial _{k} \]
Small-scale oscillations of the magnetic field are easily found from the equation (\ref{eqII2}):
\begin{equation}\label{eqII4}
  B_{0}^{i} =\frac{\left(H_{p} \partial _{p} \right)}{\widehat{D}_{H}
} v_{0}^{i}
\end{equation}
We substitute (\ref{eqII4}) into (\ref{eqII1}),  then, using the condition of solenoidality fields  $\vec{v}_{0} $  and   $\vec{B}_{0} $  (\ref{eqII3}), we find the pressure $P_{0}$:
\begin{equation} \label{eqII5}
 P_{0} =\widehat{P}_{1} u_{0} +\widehat{P}_{2}
 v_{0} +\widehat{P}_{3} w_{0}
\end{equation}
Here we introduced the designation for operators
\[\widehat{P}_{1} =\frac{D_{2} \partial _{z}
-D_{3} \partial _{y} }{\partial ^{2} } -\widetilde{Q}H_{1} \frac{\left(H_{p} \partial
_{p} \right)}{\widehat{D}_{H} } ,\]
\begin{equation} \label{eqII6}
 \widehat{P}_{2} =\frac{D_{3} \partial _{x} -D_{1}
\partial _{z} }{\partial ^{2} } -\widetilde{Q}H_{2} \frac{\left(H_{p} \partial _{p}
\right)}{\widehat{D}_{H} } ,\quad \widehat{P}_{3} =\frac{D_{1} \partial _{y} -D_{2} \partial
_{x} }{\partial ^{2} }
\end{equation}
and velocities:$v_{0}^{x} =u_{0} $, $v_{0}^{y} =v_{0} $, $v_{0}^{z} =w_{0} $. Using the representation (\ref{eqII5}), we can eliminate pressure from the equations (\ref{eqII1}) and obtain equation system for velocity fields of the zero approximation:
\[\left(\widehat{D}_{W} -\frac{\widetilde{Q}\left(H_{p} \partial _{p} \right)^{2}
}{\widehat{D}_{H} } +\widehat{p}_{1x} \right)u_{0} +\left(\widehat{p}_{2x} -D_{3}
\right)v_{0} +\left(\widehat{p}_{3x} +D_{2} \right)w_{0} =F_{0}^{x} \]

\begin{equation}\label{eqII7}
\left(D_{3}
+\widehat{p}_{1y} \right)u_{0} +\left(\widehat{D}_{W} -\frac{\widetilde{Q}\left(H_{p}
\partial _{p} \right)^{2} }{\widehat{D}_{H} } +\widehat{p}_{2y} \right)v_{0} +\left(
\widehat{p}_{3y} -D_{1} \right)w_{0} =F_{0}^{y}
\end{equation}

  \[\left(\widehat{p}_{1z} -D_{2} \right)u_{0} +\left(\widehat{p}_{2z} +D_{1} \right)v_{0}
+\left(\widehat{D}_{W} -\frac{\widetilde{Q}\left(H_{p} \partial _{p} \right)^{2}
}{\widehat{D}_{H} } +\widehat{p}_{3z} \right)w_{0} =0\]
The components of the tensor  $\widehat{p}_{ij}$ have the following form:
\[\widehat{p}_{1x} =\frac{D_{2} \partial _{x}
\partial _{z} -D_{3} \partial _{x} \partial _{y} }{\partial ^{2} } ,\widehat{p}_{2x}
=\frac{D_{3} \partial ^{2} _{x} -D_{1} \partial _{x} \partial _{z} }{\partial ^{2}
} ,\widehat{p}_{3x} =\frac{D_{1} \partial _{x} \partial _{y} -D_{2} \partial ^{2}
_{x} }{\partial ^{2} } ,\]
\begin{equation} \label{eqII8}
\widehat{p}_{1y} =\frac{D_{2} \partial _{y} \partial _{z}
-D_{3} \partial ^{2} _{y} }{\partial ^{2} }  , \widehat{p}_{2y} =\frac{D_{3} \partial
_{y} \partial _{x} -D_{1} \partial _{y} \partial _{z} }{\partial ^{2} } ,\widehat{p}_{3y}
=\frac{D_{1} \partial ^{2} _{y} -D_{2} \partial _{y} \partial _{x} }{\partial ^{2}
} ,
\end{equation}
\[ \widehat{p}_{1z} =\frac{D_{2} \partial ^{2} _{z} -D_{3} \partial _{z} \partial
_{y} }{\partial ^{2} } ,\; \widehat{p}_{2z} =\frac{D_{3} \partial _{z} \partial _{x}
-D_{1} \partial ^{2} _{z} }{\partial ^{2} } , \widehat{p}_{3z} =\frac{D_{1} \partial
_{z} \partial _{y} -D_{2} \partial _{z} \partial _{x} }{\partial ^{2} }\]
The solution for equations system (\ref{eqII7}) can be found in accordance with Cramer's rule:
\[u_{0} =\frac{1}{\Delta } \left\{ \left[\left(\widehat{D}_{W} -\widetilde{Q}\frac{\left(H_{p} \partial _{p} \right)^{2}
}{\widehat{D}_{H} } +\widehat{p}_{2y} \right)\left(\widehat{D}_{W} -\frac{\widetilde{Q}
\left(H_{p} \partial _{p} \right)^{2} }{\widehat{D}_{H} } +\widehat{p}_{3z} \right)-
\left(\widehat{p}_{2z} +D_{1} \right)\left(\widehat{p}_{3y} -D_{1} \right)\right]F_{0}^{x}
\right. +  \]

\begin{equation} \label{eqII9}
 \left.{+ \left[\left(\widehat{p}_{3x} +D_{2} \right)\left(\widehat{p}_{2z} +D_{1} \right)-
\left(\widehat{p}_{2x} -D_{3} \right)\left(\widehat{D}_{W} -\frac{\widetilde{Q}\left(H_{p}
\partial _{p} \right)^{2} }{\widehat{D}_{H} } +\widehat{p}_{3z} \right) \right]F_{0}^{y}
} \right\}
\end{equation}

\[ v_{0} =\frac{1}{\Delta } \left\{ \left[\left(\widehat{D}_{W} -\widetilde{Q}\frac{\left(H_{p} \partial _{p} \right)^{2}
}{\widehat{D}_{H} } -\widehat{p}_{1x} \right)\left(\widehat{D}_{W} -\frac{\widetilde{Q}
\left(H_{p} \partial _{p} \right)^{2} }{\widehat{D}_{H} } +\widehat{p}_{3z} \right)-
\left(\widehat{p}_{3x} +D_{2} \right)\left(\widehat{p}_{1z} -D_{2} \right) \right]F_{0}^{y}
+ \right.\]

\begin{equation} \label{eqII10} \left. {+\left[\left(\widehat{p}_{3y} -D_{1} \right) \left(\widehat{p}_{1z} -D_{2} \right)-
\left(D_{3} +\widehat{p}_{1y} \right)\left(\widehat{D}_{W} -\frac{\widetilde{Q}\left(H_{p}
\partial _{p} \right)^{2} }{\widehat{D}_{H} } +\widehat{p}_{3z} \right) \right]F_{0}^{x}
} \right\} \end{equation}

\[ w_{0} =\frac{1}{\Delta } \left\{ \left[\left(D_{3} +\widehat{p}_{1y} \right)\left(\widehat{p}_{2z} +D_{1} \right)-\left(
\widehat{D}_{W} -\frac{\widetilde{Q}\left(H_{p} \partial _{p} \right)^{2} }{\widehat{D}_{H}
} +\widehat{p}_{2y} \right)\left(\widehat{p}_{1z} -D_{2} \right) \right]F_{0}^{x} + \right.\]

\begin{equation} \label{eqII11}
\left.{+\left[\left(\widehat{p}_{2x} -D_{3} \right)\left(\widehat{p}_{1z} -D_{2} \right)-
\left(\widehat{D}_{W} -\widetilde{Q}\frac{\left(H_{p} \partial _{p} \right)^{2} }{
\widehat{D}_{H} } +\widehat{p}_{1x} \right)\left(\widehat{p}_{2z} +D_{1} \right)
\right]F_{0}^{y} }\right\}
\end{equation}
Here  $\Delta$  is the determinant of equation system (\ref{eqII7}):
\[\Delta =\left(\widehat{D}_{W}
-\widetilde{Q}\frac{\left(H_{p} \partial _{p} \right)^{2} }{\widehat{D}_{H} } +\widehat{p}_{1x}
\right)\left(\widehat{D}_{W} -\widetilde{Q}\frac{\left(H_{p} \partial _{p} \right)^{2}
}{\widehat{D}_{H} } +\widehat{p}_{2y} \right)\left(\widehat{D}_{W} -\frac{\widetilde{Q}
\left(H_{p} \partial _{p} \right)^{2} }{\widehat{D}_{H} } +\widehat{p}_{3z} \right)+\]

\[+\left(D_{3} +\widehat{p}_{1y} \right)\left(\widehat{p}_{2z} +D_{1} \right)\left(
\widehat{p}_{3x} +D_{2} \right)+\left(\widehat{p}_{2x} -D_{3} \right)\left(\widehat{p}_{3y}
-D_{1} \right)\left(\widehat{p}_{1z} -D_{2} \right)-\]

\[-\left(\widehat{p}_{3x}
+D_{2} \right)\left(\widehat{D}_{W} -\frac{\widetilde{Q}\left(H_{p} \partial _{p}
\right)^{2} }{\widehat{D}_{H} } +\widehat{p}_{2y} \right)\left(\widehat{p}_{1z} -D_{2}
\right)-\]

\[-\left(\widehat{p}_{2z} +D_{1} \right)\left(\widehat{p}_{3y} -D_{1}
\right)\left(\widehat{D}_{W} -\frac{\widetilde{Q}\left(H_{p} \partial _{p} \right)^{2}
}{\widehat{D}_{H} } +\widehat{p}_{1x} \right)-\]

\begin{equation}\label{eqII12}
    -\left(D_{3} +\widehat{p}_{1y}
\right)\left(\widehat{p}_{2x} -D_{3} \right)\left(\widehat{D}_{W} -\frac{\widetilde{Q}
\left(H_{p} \partial _{p} \right)^{2} }{\widehat{D}_{H} } +\widehat{p}_{3z} \right)
\end{equation}
In order to calculate the expressions (\ref{eqII9})-(\ref{eqII12}) we present the external force (\ref{eq6})  in complex form:
\begin{equation} \label{eqII13}
 \vec{F}_{0} = \vec{i} \, \frac{f_{0} }{2} \; e^{i
\phi _{2} } +\vec{j} \, \frac{f_{0} }{2} e^{i\phi _{1} } + k.c.
\end{equation}
Then all operators in formulae (\ref{eqII9})-(\ref{eqII12}) act from the left on their eigenfunction. In particular:
\[\widehat{D}_{W,H} e^{i\phi _{1} } =e^{i\phi
_{1} } \widehat{D}_{W,H} \left(\vec{\kappa }_{1} , -\omega _{0} \right), \quad \widehat{D}_{W,H}
e^{i\phi _{2} } =e^{i\phi _{2} } \widehat{D}_{W,H} \left(\vec{\kappa }_{2} ,- \omega _{0} \right), \]
\begin{equation} \label{eqII14}
 \Delta e^{i\phi _{1} } =e^{i\phi _{1} } \Delta \left(\vec{\kappa
}_{1} ,\; -\omega _{0} \right),\quad \Delta e^{i\phi _{2} } =e^{i\phi _{2} } \Delta \left(
\vec{\kappa }_{2} ,\; -\omega _{0} \right)
\end{equation}
To simplify the formulae, let us choose  $\kappa _{0} =1$,  $\omega _{0} =1$  and introduce new designations:
\begin{equation}\label{eqII15}
\widehat{D}_{W} \left(\vec{\kappa }_{1} ,\; -\omega _{0} \right)=
\widehat{D}_{W_{1} }^{*} =1-i\left(1-W_{1} \right),\quad \widehat{D}_{W} \left(\vec{\kappa
}_{2} ,\; -\omega _{0} \right)=\widehat{D}_{W_{2} }^{*} =1-i\left(1-W_{2} \right) \end{equation}

\[\widehat{D}_{H}
\left(\vec{\kappa }_{1} ,\; -\omega _{0} \right)=\widehat{D}_{H_{1} }^{*} =Pm^{-1}
-i\left(1-W_{1} \right),\quad \widehat{D}_{H} \left(\vec{\kappa }_{2} ,\; -\omega _{0}
\right)=\widehat{D}_{H_{2} }^{*} =Pm^{-1} -i\left(1-W_{2} \right)\]
Complex-conjugate quantities  will be denoted with asterisk. When performing further calculations, the part of component tensors  $\widehat{p}_{ij} \left(\vec{\kappa}_{1} \right)$ and  $\widehat{p}_{ij} \left(\vec{\kappa }_{2} \right)$ vanishes. Taking this into account the velocity field of  zero approximation has the following form:
\begin{equation}\label{eqII16}
 u_{0} =\frac{f_{0} }{2} \frac{A_{2} }{A_{2}^{2} +D_{2}^{2} }
e^{i\phi _{2} } +c.c.=u_{03} +u_{04}
\end{equation}

\begin{equation} \label{eqII17}
 v_{0} =\frac{f_{0} }{2} \frac{A_{1} }{A_{1}^{2}
+D_{1}^{2} } e^{i\phi _{1} } +c.c.=v_{01} +v_{02}
\end{equation}

\begin{equation} \label{eqII18}
 w_{0} =-\frac{f_{0} }{2} \frac{D_{1} }{A_{1}^{2}
+D_{1}^{2} } e^{i\phi _{1} } +\frac{f_{0} }{2} \frac{D_{2} }{A_{2}^{2} +D_{2}^{2}
} e^{i\phi _{2} } +c.c.=w_{01} +w_{02} +w_{03} +w_{04}
\end{equation}
where \[A_{1,2} =\widehat{D}_{W_{1,2} }^{*} +\widetilde{Q}\frac{H_{1,2}^{2} }{\widehat{D}_{H_{1,2}
}^{*} } \]
Components of velocity satisfy the following relations: $w_{02} =\left(w_{01}
\right)^{*} $, $w_{04} =\left(w_{03} \right)^{*} $, $v_{02} =\left(v_{01} \right)^{*}
, v_{04} =\left(v_{03} \right)^{*} $, $ u_{02} =\left(u_{01} \right)^{*} $, $u_{04} =\left(u_{03} \right)^{*}$.
In the limiting case $\sigma = 0$,  the non-conductive fluid of the formula (\ref{eqII16})-(\ref{eqII18}) coincide with the results of  \cite{41s}.  We turn now to the computation of small-scale oscillations of the field $\vec{B}_{0}$ , using the expression (\ref{eqII4}) and (\ref{eqII16})-(\ref{eqII18}):
\begin{equation} \label{eqII19}
 B_{0}^{x} =\widetilde{u}_{0} =\frac{f_{0} }{2}
\left(\frac{A_{2} }{A_{2}^{2} +D_{2}^{2} } \right)\frac{iH_{2} }{\widehat{D}_{H_{2}
}^{*} } e^{i\phi _{2} } +c.c.=\widetilde{u}_{03} +\widetilde{u}_{04}
\end{equation}

\begin{equation}\label{eqII20}
 B_{0}^{y} =\widetilde{v}_{0} =\frac{f_{0} }{2} \left(\frac{A_{1}
}{A_{1}^{2} +D_{1}^{2} } \right)\frac{iH_{1} }{\widehat{D}_{H_{1} }^{*} } e^{i\phi
_{1} } +c.c.=\widetilde{v}_{01} +\widetilde{v}_{02}  \end{equation}

\[B_{0}^{z} =\widetilde{w}_{0} =-\frac{f_{0}
}{2} \left(\frac{D_{1} }{A_{1}^{2} +D_{1}^{2} } \right)\frac{iH_{1} }{\widehat{D}_{H_{1}
}^{*} } e^{i\phi _{1} } +\frac{f_{0} }{2} \left(\frac{D_{2} }{A_{2}^{2} +D_{2}^{2}
} \right)\frac{iH_{2} }{\widehat{D}_{H_{2} }^{*} } e^{i\phi _{2} } +c.c. =\]

\begin{equation} \label{eqII21}
 =\widetilde{w}_{01}
+\widetilde{w}_{02} +\widetilde{w}_{03} +\widetilde{w}_{04}
\end{equation}
Let us note that, in the otained here expressions,  for small-scale oscillations of the velocity field  $\vec{v}_{0} $  and the magnetic field $\vec{B}_{0}$, the component of the angular velocity $D_{3} $  has been dropped out due to the choice of the external force. Next, the results (\ref{eqII16})-(\ref{eqII21}) will be used to calculate the correlation functions.

\section*{Appendix III. Calculation of the Reynolds stresses, Maxwell stresses and turbulent e.m.f.}

To close the equations (\ref{eq17})-(\ref{eq20}) we have to calculate  correlators of  the following form:

\begin{equation}\label{eqIII1}
 T^{31} =\overline{w_{0} u_{0} }=\overline{w_{01} \left(u_{01}
\right)^{*} }+\overline{\left(w_{01} \right)^{*} u_{01} }+\overline{w_{03} \left(u_{03}
\right)^{*} }+\overline{\left(w_{03} \right)^{*} u_{03} }
\end{equation}

\begin{equation} \label{eqIII2}
 T^{32} =\overline{w_{0} v_{0} }=\overline{w_{01}
\left(v_{01} \right)^{*} }+\overline{\left(w_{01} \right)^{*} v_{01} }+\overline{w_{03}
\left(v_{03} \right)^{*} }+\overline{\left(w_{03} \right)^{*} v_{03} }
\end{equation}

\begin{equation}\label{eqIII3}
 S^{31} =\overline{\widetilde{w}_{0} \widetilde{u}_{0} }=\overline{
\widetilde{w}_{01} \left(\widetilde{u}_{01} \right)^{*} }+\overline{\left(\widetilde{w}_{01}
\right)^{*} \widetilde{u}_{01} }+\overline{\widetilde{w}_{03} \left(\widetilde{u}_{03}
\right)^{*} }+\overline{\left(\widetilde{w}_{03} \right)^{*} \widetilde{u}_{03} }
\end{equation}

\begin{equation}\label{eqIII4}
 S^{32} =\overline{\widetilde{w}_{0} \widetilde{v}_{0} }=\overline{
\widetilde{w}_{01} \left(\widetilde{v}_{01} \right)^{*} }+\overline{\left(\widetilde{w}_{01}
\right)^{*} \widetilde{v}_{01} }+\overline{\widetilde{w}_{03} \left(\widetilde{v}_{03}
\right)^{*} }+\overline{\left(\widetilde{w}_{03} \right)^{*} \widetilde{v}_{03} }
\end{equation}

\begin{equation}\label{eqIII5}
 G^{13} =\overline{u_{0} \widetilde{w}_{0} }=\overline{u_{01}
\left(\widetilde{w}_{01} \right)^{*} }+\overline{\left(u_{01} \right)^{*} \widetilde{w}_{01}
}+\overline{u_{03} \left(\widetilde{w}_{03} \right)^{*} }+\overline{\left(u_{03}
\right)^{*} \widetilde{w}_{03} }
\end{equation}

\begin{equation} \label{eqIII6}
 G^{31} =\overline{w_{0} \widetilde{u}_{0} }=
\overline{w_{01} \left(\widetilde{u}_{01} \right)^{*} }+\overline{\left(w_{01} \right)^{*}
\widetilde{u}_{01} }+\overline{w_{03} \left(\widetilde{u}_{03} \right)^{*} }+\overline{
\left(w_{03} \right)^{*} \widetilde{u}_{03} }
\end{equation}

\begin{equation} \label{eqIII7}
 G^{23} =\overline{v_{0} \widetilde{w}_{0} }=
\overline{v_{01} \left(\widetilde{w}_{01} \right)^{*} }+\overline{\left(v_{01} \right)^{*}
\widetilde{w}_{01} }+\overline{v_{03} \left(\widetilde{w}_{03} \right)^{*} }+\overline{
\left(v_{03} \right)^{*} \widetilde{w}_{03} }
\end{equation}

\begin{equation} \label{eqIII8}
 G^{32} =\overline{w_{0} \widetilde{v}_{0} }=
\overline{w_{01} \left(\widetilde{v}_{01} \right)^{*} }+\overline{\left(w_{01} \right)^{*}
\widetilde{v}_{01} }+\overline{w_{03} \left(\widetilde{v}_{03} \right)^{*} }+\overline{
\left(w_{03} \right)^{*} \widetilde{v}_{03} }
\end{equation}
We begin with the calculations of the Reynolds stresses determined by the formulas (\ref{eqIII1}),
(\ref{eqIII2}). First of all, we need expressions for small-scale velocity fields (\ref{eqII16})-(\ref{eqII18}). After substituting them  into the formulas (\ref{eqIII1}), (\ref{eqIII2}) we get:
\begin{equation} \label{eqIII9}
 T^{31} =\frac{f_{0}^{2} }{2} \frac{D_{2} q_{2}
}{\left|A_{2}^{2} +D_{2}^{2} \right|^{2} }
\end{equation}

\begin{equation}\label{eqIII10}
    T^{32} =-\frac{f_{0}^{2} }{2} \frac{D_{1} q_{1} }{\left|A_{1}^{2} +D_{1}^{2} \right|^{2}}
\end{equation}
where  \[q_{1,2} =1 + \frac{Q H_{1,2}^{2}}{1+Pm^{2} \left(1-W_{1,2} \right)^{2}}\]
Now we proceed to calculate the magnetic field correlators or Maxwell stresses  $S^{31} $ and $S^{32} $, using the formulas (\ref{eqII19})-(\ref{eqII21}). Then, as a result of the substitution (\ref{eqII19})-(\ref{eqII21}) in (\ref{eqIII3}), (\ref{eqIII4}) we find:
\begin{equation}\label{eqIII11}
    S^{31} =\frac{H_{2}^{2} }{\left|\widehat{D}_{H_{2} } \right|^{2}
} T^{31}, \quad S^{32} =\frac{H_{1}^{2} }{\left|\widehat{D}_{H_{1}
} \right|^{2} } T^{32}
\end{equation}
Since the right-hand sides of the equations (\ref{eq17}), (\ref{eq18}) contain the difference  $T^{31} -\widetilde{Q}S^{31}$  and     $T^{32} -\widetilde{Q}S^{32} $, it is easy to find it using the expressions (\ref{eqIII10})-(\ref{eqIII11}):
\begin{equation} \label{eqIII12}
 T^{31} -\widetilde{Q}S^{31} =T^{31} \left(1-
\frac{QH_{2}^{2} }{Pm\left|\widehat{D}_{H_{2} } \right|^{2} } \right)=T^{31} Q_{2}
\end{equation}

\begin{equation}\label{eqIII13}
 T^{32} -\widetilde{Q}S^{32} =T^{32} \left(1-\frac{QH_{1}^{2}
}{Pm\left|\widehat{D}_{H_{1} } \right|^{2} } \right)=T^{32} Q_{1}
\end{equation}
Taking into account the relations (\ref{eqIII11}) and opening modules in formulas(\ref{eqIII12}),
(\ref{eqIII13}) , we obtain expressions  in the form:
\begin{equation} \label{eqIII14}
 T^{31} -\widetilde{Q}S^{31} =\frac{f_{0}^{2}
}{2} \frac{D_{2} q_{2} Q_{2} }{\left[4\left(1-W_{2} \right)^{2} q_{2}^{2} Q_{2}^{2}
+\left[D_{2}^{2} +W_{2} \left(2-W_{2} \right)+\mu _{2} \right]^{2} \right]}
\end{equation}

\begin{equation}\label{eqIII15}
 T^{32} -\widetilde{Q}S^{32} =-\frac{f_{0}^{2} }{2} \frac{D_{1}
q_{1} Q_{1} }{\left[4\left(1-W_{1} \right)^{2} q_{1}^{2} Q_{1}^{2} +\left[D_{1}^{2}
+W_{1} \left(2-W_{1} \right)+\mu _{1} \right]^{2} \right]}
\end{equation}

\[\mu _{1,2} =(q_{1,2} -1)[2(1+Pm\left(1-W_{1,2}^{2} \right)^{2} +(q_{1,2} -1)(1-Pm^{2}
\left(1-W_{1,2}^{2} \right)^{2} ]\]
Now we turn to the computation of correlators determined by the formulas (\ref{eqIII5})-(\ref{eqIII8}). In order to do  this, we need expressions for the small-scale velocity field (\ref{eqII16})-(\ref{eqII18}) and magnetic field (\ref{eqII19})-(\ref{eqII21}). As a result of simple mathematical operations, we find:
\begin{equation} \label{eqIII16}
 G^{13} =\frac{f_{0}^{2} }{4} \frac{iH_{2} D_{2}
}{\left|A_{2}^{2} +D_{2}^{2} \right|^{2} } \left(\frac{A_{2}^{*} }{\widehat{D}_{H_{2}
}^{*} } -\frac{A_{2} }{\widehat{D}_{H_{2} } } \right)
\end{equation}

\begin{equation} \label{eqIII17}
 G^{31} =\frac{f_{0}^{2} }{4} \frac{iH_{2} D_{2}
}{\left|A_{2}^{2} +D_{2}^{2} \right|^{2} } \left(\frac{A_{2} }{\widehat{D}_{H_{2}
}^{*} } -\frac{A_{2} ^{*} }{\widehat{D}_{H_{2} } } \right)
\end{equation}

\begin{equation} \label{eqIII18}
 G^{23} =\frac{f_{0}^{2} }{4} \frac{iH_{1} D_{1}
}{\left|A_{1}^{2} +D_{1}^{2} \right|^{2} } \left(\frac{A_{1} }{\widehat{D}_{H_{1}
} } -\frac{A_{1}^{*} }{\widehat{D}_{H_{1} }^{*} } \right)
\end{equation}

\begin{equation} \label{eqIII19}
 G^{32} =\frac{f_{0}^{2} }{4} \frac{iH_{1} D_{1}
}{\left|A_{1}^{2} +D_{1}^{2} \right|^{2} } \left(\frac{A_{1} ^{*} }{\widehat{D}_{H_{1}
} } -\frac{A_{1} }{\widehat{D}_{H_{1} }^{*} } \right)
\end{equation}
To close the large-scale magnetic field equations (\ref{eq19}), (\ref{eq20}), we need to calculate the differences $G^{13} -G^{31} $,  $G^{23} -G^{32} $, which correspond to the components of a turbulent e.m.f. ${\mathcal{E}}_{2}={\mathcal{E}}_{y}$ and ${\mathcal{E}}_{1}={\mathcal{E}}_{x}$. Taking into account the formulas  (\ref{eqIII16})-(\ref{eqIII19})  we obtain:
\begin{equation} \label{eqIII20}
 {\mathcal{E}}_{2}=G^{13} -G^{31} =
\frac{f_{0}^{2} }{4} \frac{iH_{2} D_{2} }{\left|A_{2}^{2} +D_{2}^{2} \right|^{2}
} \frac{(A_{2}^{*} -A_{2} )(\widehat{D}_{H_{2} }^{*} +\widehat{D}_{H_{2} } )}{\left|
\widehat{D}_{H_{2} } \right|^{2} }
\end{equation}

\begin{equation} \label{eqIII21}
 {\mathcal{E}}_{1}  =G^{23} -G^{32} =-
\frac{f_{0}^{2} }{4} \frac{iH_{1} D_{1} }{\left|A_{1}^{2} +D_{1}^{2} \right|^{2}
} \frac{(A_{1}^{*} -A_{1} )(\widehat{D}_{H_{1} }^{*} +\widehat{D}_{H_{1} } )}{\left|
\widehat{D}_{H_{1} } \right|^{2} }
\end{equation}
Using    $A_{1,2}$, $\widehat{D}_{H_{1,2}}$   from the formulas (\ref{eqIII20}), (\ref{eqIII21}) we find the difference of the mixed correlators or the components of the turbulent e.m.f. in an explicit form:
\begin{equation} \label{eqIII22} {\mathcal{E}}_{2} =-f_{0}^{2} \frac{D_{2}
\left(1-W_{2} \right)PmQ_{2} H_{2} }{\left(1+Pm^{2} \left(1-W_{2} \right)^{2} \right)
\left[4\left(1-W_{2} \right)^{2} q_{2}^{2} Q_{2}^{2} +\left[D_{2}^{2} +W_{2} \left(2-W_{2}
\right)+\mu _{2} \right]^{2} \right]}  \end{equation}
\begin{equation}\label{eqIII23} {\mathcal{E}}_{1} =f_{0}^{2} \frac{D_{1} \left(1-W_{1}
\right)PmQ_{1} H_{1} }{\left(1+Pm^{2} \left(1-W_{1} \right)^{2} \right)\left[4\left(1-W_{1}
\right)^{2} q_{1}^{2} Q_{1}^{2} +\left[D_{1}^{2} +W_{1} \left(2-W_{1} \right)+\mu
_{1} \right]^{2} \right]}  \end{equation}


\begin{thebibliography}{44}

\bibitem{1s} G. Moffat, Magnetic Field Generation in Electrically Conducting Fluids (Cambridge University Press, Cambridge, 1978; Mir, Moscow, 1980).
\bibitem{2s} Ya.B. Zeldovich, A.A. Ruzmaykin and D.D. Sokoloff, Magnetic Fields in Astrophysics (RHD, Izhevsk, 2006) [in Russian].
\bibitem{3s} S.I. Vainshtein, Ya.B. Zeldovich and A.A. Ruzmaikin, Turbulent Dynamo in Astrophysics (Nauka, Moscow,1980) [in Russian].
\bibitem{4s} S.I. Vainshtein, Magnetic Fields in Space ( Nauka, Moscow, 1983) [in Russian].
\bibitem{5s} J. Parker, Conversations on Electric and Magnetic Fields in the Cosmos (Princeton University Press, Princeton, 2007).
\bibitem{6s} F. Krause, K-H. R\"{a}dler, Mean-Field Magnetohydrodynamics and Dynamo Theory (Academie-Verlag, Berlin, 1990).
\bibitem{7s} Ruzmajkin A.A., Sokolov D.D., Shukurov A.M., Magnetic fields of Galaxies ( Nauka, Moscow,  1988)  [in Russian].
\bibitem{8s} D.D. Sokoloff, R.A. Stepanov and P.G. Frick, Phys.-Usp. 184, 318 (2014).
\bibitem{9s} V.I Arnold, Ya.B.., Zeldovich ,  A.A.Ruzmajkin ,  D.D. Sokolov, Magnetic Field in a Moving Conducting Fluid, Uspehi matem. Nauk,  No5, P.220-221  (1981).
\bibitem{10s} V.I. Arnold, Evolution of Magnetic Field under the Action of Transfer and Diffusion, Uspehi matem. Nauk,  No. 2, P.225-227  (1983).
\bibitem{11s} H.  Grinspen, The Theory of Rotating Fluids ( Gidrometeoizdat, Leningrad, 1975) [in Russian].
\bibitem{12s} Rotating Fluids in Geophysics.  P.H.Roberts and A.M.Soward eds., Acad.Press (1978).
\bibitem{13s} Dj. Pedloski,  Geophysical Fluid Dynamics ( Mir, Moscow, 1984)  [in Russian].
\bibitem{14s} V.I.  Petviashvili , O.A.  Pohotelov, Solitary waves in Plasma and Atmosphere (Energoatomisdat, , Moscow, 1989) [in Russian].
\bibitem{14ss} G.D.  Aburdjanija, Self-organization of Nonlinear Vortex Structures and Vortex Turbulence in Dispersive Media (KomKniga, Moscow, 2006) [in Russian].
\bibitem{16s}  A.V. Kolesnichenko, M.Ya. Marov, Turbulence and self-organization. Problems of modeling space and natural environments  (BINOM, Moscow, 2009) [in Russian].
\bibitem{17s} O.G. Onishenko, O.A. Pohotelov ,  N.M. Astafjeva, Usp. Fis. Nauk., Vol.178, No. 6, P. 605-618  (2008).
\bibitem{18s} M.V. Nezlin, E.N. Sneshkin, Rossby Vortices and Spiral Structure (Nauka, Moscow, 1990) [in Russian].
\bibitem{19s} Moffatt H.K. The degree of knottedness of tangled vortex lines. J. Fluid Mech. 1969. V.35. P.117-129.
\bibitem{20s} M.  Steenbek, F. Krause,  Magnitnaja gidrodinamika,  No.3.  P. 19-44  ( 1967)  [in Russian].
\bibitem{21s} H.K. Moffat, Some directions of development of turbulence. Modern fluid dynamics. Successes and problems. ( Mir, Moscow, 1984)  [in Russian].
\bibitem{22s} F. Krause, G. R\"{u}diger , Astron. Nachr., V. 295, P.93-99 (1974).				
\bibitem{23s} S.S. Moiseev, R.Z. Sagdeev, A.V. Tur, G.A Homenko, V.V. Yanovsky, Sov. Phys. JETP,  85, 1979-1987  (1983).
\bibitem{24s} S.S.  Moiseev, P.B. Rutkevitch,  A.V. Tur,  V.V.  Yanovsky, Sov.Phys.  JETP, 67, 294 (1988).
\bibitem{25s} E.A. Lypyan, A.A.  Mazurov ,   P.B.  Rutkevitch, A.V. Tur,  Sov.Phys.  JETP, 75, 838 (1992).
\bibitem{26s} S.S. Moiseev, K.R. Oganjan, P.B.  Rutkevith,  A.V. Tur, G.A. Homenko,  V.V. Yanovsky, Vortex Dynamo in Helical Turbulence ( Naukova dumka, Kiev , 1990) [in Russian].
\bibitem{27s} Levina G.V., Moiseev S.S., Rutkevitch P.B. Hydrodynamic alpha-effect in a convective
system. Advance in Fluid Mechanics, 25, 111 (2000)
\bibitem{28s} V.D. Zimin, G.V. Levina, S.S. Moiseev, A.V. Tur. Dokl. AN SSSR, Vol. 309, P. 88-92 (1989).
\bibitem{29s} Tur A.V., Yanovsky V.V. Large-scale instability in hydrodynamics with stable
temperature stratification driven by small-scale helical force.ar Xiv:1204.5024 v.1[physics.
Flu-dyn.](2012)
\bibitem{30s} A.V. Tur , V.V. Yanovsky  Non Linear Vortex Structure in Stratified Driven by Small-scale
Helical Forse.Open Journal of Fluid Dynamics, 3, 64-74 (2013)
\bibitem{31s} M.I. Kopp, A.V. Tur, V.V. and Yanovsky, JETP, 147(4), 846 (2015).
\bibitem{32s}  P. B. Rutkevich,  Equation for vortex Instability Caused by Convective Turbulence
and Coriolis Force, JETF, 1993, v. 77, p.933.
\bibitem{33s} L. M.  Smith and  F. Waleffe, Transfer of Energy to Two-Dimensional Large Scales
in Forced, Rotating Three-Dimensional Turbulence, Physics of Fluids, 1999, v. 11,
No. 6, p.1608.
\bibitem{34s} L. M. Smith and F. Waleffe,  Generation of Slow Large Scales in Forced Rotating
Stratified Turbulence, Journal of Fluid Mechanics, 2002, v.451, pp. 145-168.
\bibitem{35s} Berezin Y. A. and Zhukov V. P. An Influence of Rotation on Convective Stability
of Large Scale Distorbances in Turbulent Fluid, Izv. AN SSSR, Mech. Zhidk. Gaza,
1989, v. 4, p.3.
\bibitem{36s} L.L. Kitchatinov, G. R\"{u}diger, and G. Khomenko, Large-scale vortices in rotating
stratified disks, Astron. Astrophys. 287,320 (1994).
\bibitem{37s} U. Frishe, Z.S. She, P.L. Sulem.  Large Scale Flow Driven by the Anisotropic Kinetic
Alpha Effect, Physica D, 28, 382 (1987).
\bibitem{38s} O.A. Druzhinin, G.A. Homenko, Nonlinear Theory of Hydrodynamic Alpha-Effect in the Compressible Fluid and the inverse Energy Cascade, Proceeding of the IV Intern.Workshop on Nonlinear and Turb. Pros.in Physics, Kiev, 1989.
\bibitem{39s}  P.B. Rutkevitch , R.Z. Sagdeev, A.V. Tur, V.V. Yanovsky.  Nonlinear dynamic theory
of the $\alpha $-effect in  compressible fluid. Proceeding of the IV  Intern. Workshop
on Nonlinear and  Turb. Pros.in Physics, Kiev, 1989.
\bibitem{40s} O.G. Chkhetiani, S.S. Moiseev, E.I. Golbraikh, Sov. Phys. JETP 87(3), 513 (1998).
\bibitem{41s} M. I. Kopp, A. V. Tur, V. V. Yanovsky. Nonlinear Vortex Structures in Obliquely Rotating Fluid. Open Journal of Fluid Dynamics, 5, 311-321 (2015)
\bibitem{42s} A.Z. Dolginov,  V.A. Urpin, Sov. Phys. JETP 77, 1921 (1978).
\bibitem{43s} G.  R\"{u}diger, On the $\alpha $- Effect for Slow and Fast Rotation, Astron. Nachr., 1978, v.299, No.4, pp. 217-222.
\bibitem{44s} Yu. A.  Danilov, Lectures on nonlinear dynamics ( KomKniga, Moscow,  2006)  [in Russian].

\end{thebibliography}
\end{document}